\documentclass[useAMS,usenatbib]{mn2e}
\usepackage[T1]{fontenc}
\usepackage{ae,aecompl}

\usepackage{graphicx}
\usepackage{tablefootnote}
\usepackage{amssymb}
\usepackage{amsmath}
\usepackage{booktabs}
\usepackage[miktex]{gnuplottex}
\usepackage{natbib}
\usepackage{caption}
\usepackage{textcomp}

\newcommand{\apj}{ApJ}
\newcommand{\apjl}{ApJ}

\newcommand{\aap}{A\&A}
\newcommand{\araa}{Annual Review of Astronomy and Astrophysics}
\newcommand{\mnras}{MNRAS}

\newcommand{\prd}{Phys. Rev. D}

\newcommand{\sjetp}{Sov. Phys.~JETP}
\newcommand{\ssr}{Space Science Reviews}
\newcommand{\rpp}{Reports on Progress in Physics}

\newcommand{\znta}{Zeitschrift Naturforschung Teil A}
\newcommand{\zap}{Z. Astrophys.}

\usepackage{hyperref}
\hypersetup{
    bookmarks=true,
    unicode=false,
    pdftoolbar=true,
    pdfmenubar=true,
    pdffitwindow=false,
    pdfstartview={FitH},
    pdftitle={My title},
    pdfauthor={Author},
    pdfsubject={Subject},
    pdfcreator={Creator},
    pdfproducer={Producer},
    pdfkeywords={keyword1} {key2} {key3},
    pdfnewwindow=true,
    colorlinks=true,
    linkcolor=red,
    citecolor=blue,
    filecolor=magenta,
    urlcolor=cyan
}

\makeatletter

\patchcmd{\NAT@citex}
  {\@citea\NAT@hyper@{%
     \NAT@nmfmt{\NAT@nm}%
     \hyper@natlinkbreak{\NAT@aysep\NAT@spacechar}{\@citeb\@extra@b@citeb}%
     \NAT@date}}
  {\@citea\NAT@nmfmt{\NAT@nm}%
   \NAT@aysep\NAT@spacechar\NAT@hyper@{\NAT@date}}{}{}

\patchcmd{\NAT@citex}
  {\@citea\NAT@hyper@{%
     \NAT@nmfmt{\NAT@nm}%
     \hyper@natlinkbreak{\NAT@spacechar\NAT@@open\if*#1*\else#1\NAT@spacechar\fi}%
       {\@citeb\@extra@b@citeb}%
     \NAT@date}}
  {\@citea\NAT@nmfmt{\NAT@nm}%
   \NAT@spacechar\NAT@@open\if*#1*\else#1\NAT@spacechar\fi\NAT@hyper@{\NAT@date}}
  {}{}

\makeatother

\title[Rotation, turbulence and Pop. III star formation]{Formation sites of Population~III star formation: The effects of different levels of rotation and turbulence on the fragmentation behavior of primordial gas} 
\author[Wollenberg et al.]{Katharina M. J. Wollenberg$^{1,2,3}$\thanks{E-mail: 
fi432@uni-heidelberg.de}, Simon C. O. Glover$^{1}$, Paul C. Clark$^{4}$, 
\newauthor{Ralf S. Klessen$^{1,5}$} \\
$^{1}$Zentrum f\"{u}r Astronomie, Universit\"{a}t Heidelberg, Institut f\"{u}r Theoretische Astrophysik,  Albert-Ueberle-Str. 2, 69120 Heidelberg, Germany\\
$^{2}$International Max Planck Research School for Astronomy \& Cosmic Physics at the University of Heidelberg\\
$^{3}$Heidelberg Graduate School of Fundamental Physics at University of Heidelberg\\
$^{4}$School of Physics and Astronomy, Queen's Buildings, The Parade, Cardiff University, Cardiff, CF24 3AA\\
$^{5}$Interdisziplin\"{a}res Zentrum fur Wissenschaftliches Rechnen, INF 205, D-69120, Heidelberg, Germany}

\begin{document}

\date{}

\pagerange{\pageref{firstpage}--\pageref{lastpage}} \pubyear{2019}

\maketitle

\label{firstpage}

\begin{abstract}
We use the moving-mesh code \textsc{arepo} to investigate the effects of different levels of rotation and turbulence on the fragmentation of primordial gas and the formation of Population III stars. We consider 9 different combinations of turbulence and rotation and carry out 5 different realizations of each setup, yielding one of the largest sets of simulations of Population III star formation ever performed.  We find that fragmentation in Population III star-forming systems is a highly chaotic process and show that the outcomes of individual realizations of the same initial conditions often vary significantly. However, some general trends are apparent. Increasing the turbulent energy promotes fragmentation, while increasing the rotational energy inhibits fragmentation.  Within the $\sim 1000$~yr period that we simulate, runs including turbulence yield flat protostellar mass functions while purely rotational runs show a more top-heavy distribution. The masses of the individual protostars are distributed over a wide range from a few $10^{-3} \, {\rm M_{\odot}}$  to several tens of ${\rm M_\odot}$. The total mass growth rate of the stellar systems remains high throughout the simulations and depends only weakly on the degree of rotation and turbulence. Mergers between protostars are common, but predictions of the merger fraction are highly sensitive to the criterion used to decide whether two protostars should merge.  Previous studies of Population~III star formation have often considered only one realization per set of initial conditions. However, our results demonstrate that robust trends can only be reliably identified by considering averages over a larger sample of runs.
\end{abstract}


\section{Introduction}
\label{Sec:Intro}
The different steps in the formation of Population III (Pop III) stars have been studied for almost two decades. At redshifts $z \sim 20-50$, primordial gas, consisting mainly of hydrogen and helium, falls into the potential wells of so-called minihalos of mass $\sim 10^6 \, \rm M_{\odot}$ and with virial temperatures $T_{\mathrm{vir}} \sim 1000 \, \rm K$ \citep{1997Tegmark,2004BrommLarson, 2005GloverReview, 2013Bromm}. During this process, the gas is heated to temperatures close to $T_{\rm vir}$, but is able to cool down afterwards via $\rm H_2$ ro-vibrational line cooling. As the gas continues to cool and condense within the minihalos, it can reach a minimum temperature of $T \sim 200 \, \rm K$ at a number density of $n \sim 10^4 \, \rm cm^{-3}$. This is comparable to the critical density of molecular hydrogen at which its rotational level populations reach local thermodynamical equilibrium (LTE). At this point of minimum temperature, the gas within a minihalo tends to break up into separate clumps of mass of the order of the local Jeans mass, $M_{\mathrm{J}} \sim 1000 \, \rm M_{\odot}$. These clumps become the birthplaces of Population III stars. 

A higher level of ionization in the gas may lead to a further elevated molecular hydrogen fraction, allowing the gas to cool to temperatures $T < 200 \, \rm K$. In this regime, the formation of deuterated hydrogen molecules, HD, is facilitated. When enough HD is created, the gas is able to cool down to even the temperature of the cosmic microwave background ($\sim 57 \, \rm K$ at $z=20$). An increased ionization fraction can for example arise when the gas in the minihalo is exposed to ionizing radiation from a nearby or progenitor Pop~III star \citep[e.g.][]{2002OhHaiman,2005Nagakura, 2008Yoshida}, to X-rays from Pop~III supernova remnants, X-ray binaries or quasars \citep[e.g.][]{2003GloverBrand, 2015Hummel}, or to cosmic rays \citep{2016Hummel}. 
The critical density in this case is $n \sim 10^6 \, \rm cm^{-3}$. The Jeans-unstable, star-forming clumps are less massive, $M_{\mathrm{J}} < 100 \, \rm M_{\odot}$, due to the smaller minimum temperature. Star formation dominated by $\rm HD$ chemistry is termed Population~III.2 in contrast to that dominated by $\rm H_2$ described above which is often  called Population~III.1 star formation \citep{2008TanMcKee}. In this study we concentrate on Population~III.1 stars, which for brevity we will generally refer to as Population~III stars.

The evolution of the star-forming gas proceeds as follows. The clumps mentioned above continue to gravitationally collapse.
 At densities of $n \sim 10^8$--$10^9 \, \rm cm^{-3}$, three-body $\rm H_2$ formation dramatically increases the amount of molecular hydrogen available in the gas and consequently increases the cooling rate. However, this rapid formation of $\rm H_2$ also converts significant amounts of chemical energy into heat, causing the temperature of the gas to increase to around 1000 -- 2000$\, \rm K$. At $n \sim 10^{10} \, \rm cm^{-3}$, the gas becomes optically thick to molecular hydrogen line cooling \citep{2004RipamontiAbel, 2006Yoshida}, but further cooling due to collision-induced emission (CIE) sets in for $n \gtrsim 10^{14} \, \rm cm^{-3}$ \citep{2004RipamontiAbel}. When the gas reaches densities of $n \gtrsim 10^{16} \, \rm cm^{-3}$, the gas also becomes optically thick to CIE cooling. At this stage, the only way in which it can dissipate the energy released during the collapse is by collisionally dissociating $\rm H_2$. This remains effective until all of the $\rm H_2$ at the center of the collapsing clump has been depleted. From there, the collapse proceeds fully adiabatically until a protostar with mass $M \lesssim 0.01 \, \rm M_{\odot}$ is born \citep{2008Yoshida}.

The newly-formed Pop III protostar is embedded within a dense, massive gas envelope, and continues to grow in mass via accretion from its surrounding gas. Since the accretion rate scales with the gas temperature as $\dot{M}_{\mathrm{acc}} \propto T^{3/2}$, typically values of $\dot{M}_{\mathrm{acc}} \gtrsim 10^{-3} \, \rm M_{\odot} \, yr^{-1}$ are encountered. Through stellar-evolution codes, it has been found that, due to these high accretion rates, pre-main sequence Population III stars are puffed-up objects with large  stellar radii, with values of up to a few hundred solar radii being common \citep{1986Stahler,2003OmukaiPalla,2009Hosokawa, 2010Hosokawa, 2012Smith, 2017Woods, 2018Haemmerle}. 

The earliest numerical studies suggested that Pop III stars form in isolation and can become very massive ($\gtrsim 100 \, \rm M_{\odot}$) \citep[see e.g.][]{2002Abel, 2002Bromm, 2004BrommLarson}. Within the last decade, however, simulations have extended the study of Pop III star formation beyond the formation of the first protostar and have shown that the non-zero angular momentum within the collapsing gas cloud leads to the formation of a self-gravitating protostellar disk that is prone to fragmentation \citep{2008Clark, 2009Turk, 2010Stacy, 2011aClark}. Instead of a single object, a protostellar cluster arises \citep{2011Greif, 2011bClark, 2011Smith, 2013StacyBromm,2017MNRAS.470..898H,Susa2019}.

The evolution of an individual protostar depends on its interactions with the surrounding gas and the other objects in the cluster. The protostars compete for further mass growth from their common mass reservoir, yielding highly variable accretion rates \citep{2011Greif, 2012Greif, 2012Girichidis, 2012Smith, 2016Hosokawa}. Some protostars might even stop accreting completely if their gas supply is removed by fragmentation or accretion onto neighboring protostellar companions, a process termed fragmentation-induced starvation in the context of present-day star formation \citep{2010Peters}. Furthermore, during close encounters protostars might get disrupted or even merge \citep{2011Greif, 2012Greif, 2016Stacy, Susa2019}. In addition, the complicated multiple-body dynamics during close interactions can lead to ejection of individual protostars from the protostellar disk or even from the halo. Indeed simulations find an ejection rate of $\sim 30 \%$ \citep{2011Greif, 2013StacyBromm, 2016Stacy}. 

Within the last couple of years simulations have tried to estimate the initial mass function (IMF) of Pop III protostellar clusters and found a top-heavy distribution in which the masses range from subsolar values to over a hundred solar masses \citep{2011Greif, 2011bClark, 2013StacyBromm, 2013Susa, 2014Hirano, 2016Stacy}. From the observational side, stellar archaeology can provide indirect constraints on the Pop III IMF through the search for metal-free stars \citep[e.g.][]{Magg2019} and for nucleosynthetic signatures of Population III stars in extremely metal-poor stars (EMP; \citealt{2005BeersChristlieb, 2015FrebelNorris}) or ultra metal-poor stars (UMP; \citealt{2013Karlsson}). 

To gain reliable estimates of the final IMF shape from numerical simulations, the evolution of the protostellar system needs to be followed through the whole accretion period. Accretion is ultimately terminated when some of the protostars have grown massive enough to produce radiative feedback that photoevaporates the accretion disk. This happens after a few $10^3$ to $10^5$ years \citep[e.g.][]{2012Stacy, 2013Susa, 2016Stacy}. 

But before we consider how a protostellar system evolves over such a long time, how massive individual stars become, and how many might be left in the end, we may ask how many protostellar objects form in the first place and how their number depends on the properties of their star-forming cloud. From earlier cosmological studies we know that the degree of fragmentation of the protostellar disk varies strongly from halo to halo \citep[e.g.][]{2011Greif, 2015aHartwig, 2013StacyBromm}. This is to be expected if turbulence within the cloud plays a role in determining when and where fragmentation actually occurs \citep[][]{2008Greif, 2011Greif}. It is also reasonable to expect that the amount of angular momentum present on small scales will have a large impact on protostellar disk formation and evolution \citep[e.g.][]{2013deSouza, 2016Stacy}. However, previous studies have only examined a very limited set of initial conditions, often drawn directly from cosmological simulation. Compared to cosmological simulations, the use of controlled initial conditions which examine the collapse of a gas cloud with pre-defined levels of turbulence or rotation within a small computational box offers several advantages. First, it is far less computationally expensive, allowing a large sample of different initial conditions to be studied. Second, it allows one to draw conclusions on the specific role, strength and importance of the particular physical parameters applied here. A few systematic studies exist of the effects of varying the initial level of turbulence \citep{2011bClark}, rotation \citep{2008Machida} or both \citep{2018Riaz} on the formation of Population III stars. However, these studies are limited because they only consider one realization per setup. From modelling of present-day star formation \citep[][]{2004aGoodwin, 2004bGoodwin}, we know that results from simulations including turbulence may vary strongly from realization to realization. This is due to the randomness of the seed with which the initial turbulent velocity field of the cloud is generated, which reflects the high degree of stochasticity observed in the turbulent interstellar medium \citep[see e.g.][]{2016KlessenGlover}. Thus, in order to be able to derive some general trends regarding the influence of rotation and turbulence, one should consider a sample of several realizations per setup.

This is what we attempt to do. In our study, we are interested in exploring how different levels of turbulence and rotation within the star-forming cloud affect the outcome of the Pop III star formation process. We pursue this examination under controlled initial conditions starting from a critical Bonnor-Ebert (BE) sphere and apply 9 setups of different combinations of rotation and turbulence with 5 realizations each, for a total of 45 simulations. 

Our paper is structured as follows. In Section~\ref{Sec:Methods}, our numerical approach with the moving mesh code \textsc{arepo} \citep{2010Springel} is explained. Hereby, we give more details on our newly implemented sink particle module and on extensions to our primordial chemistry. Section \ref{Sec:IC} outlines the initial conditions of our different setups. In Section \ref{Sec:Results}, we present the outcome of our study of the fragmentation behavior of Population III protostellar disks under the influence of different levels of rotation and turbulence. We describe the results in terms of the evolution of sink particle properties and sink dynamical behavior, including number of sinks and mass functions, accretion behavior and mergers. We discuss some shortcomings of our present study in Section \ref{sec:caveat} before we state our conclusions in Section \ref{sec:conclusion}.

\section{Numerical Method}
\label{Sec:Methods}

The studies of primordial gas cloud collapse presented in this paper are performed with an updated version of the Voronoi moving-mesh code \textsc{arepo} \citep{2010Springel} including recent improvements to the time integration scheme, spatial gradient reconstruction and grid regularization \citep[][]{2016Pakmor, 2015Mocz}. We model gas hydrodynamics using the HLLD Riemann solver \citep[][]{2005MiyoshiKusano, 2011Pakmor}. We apply the Jeans refinement criterion to make sure that the local Jeans length is always resolved by at least 16 cells, in order to avoid artificial fragmentation \citep{1997Truelove, 1997BateBurkert, 2011bFederrath}. The Voronoi mesh-generating points follow a quasi-Lagrangian behavior by being advected with the underlying gas flow. This allows the mesh to follow the growth of density fluctuations under their own self-gravity while continuously adjusting the resolution of the grid. In this way, \textsc{arepo} is an ideal code to study gas collapse \citep[][]{2010Springel, 2011Greif, 2012Greif}.

\subsection{Chemistry and cooling}
Our chemical network follows a total of 45 reactions between the twelve species $\rm H$, $\rm H^+$, $\rm H^-$, $\rm H_2^+$, $\rm H_2$, $\rm He$, $\rm He^+$, $\rm He^{++}$, $\rm D$, $\rm D^+$, $\rm HD$ and free electrons described in \citet{2011bClark}, which is based in turn on \citet{2007GloverJappsen} and \citet{2008GloverAbel}. We also use a cooling function that is based on that in \citet{2011bClark}. However, the treatment of chemistry and cooling in this study contains several improvements in the form of updated rate coefficients, as summarized in \citet{2017Schauer}. In the context of this work, the most important change is that we now use a rate coefficient for the three-body H$_{2}$ formation reaction 
\begin{equation}
{\rm H + H + H} \rightarrow {\rm H_{2} + H},
\end{equation}
computed by \citet{2013Forrey}. This yields a more reliable treatment at low gas temperatures than one obtains by applying detailed balance to the collisional dissociation reaction \citep[see e.g.][]{1983Palla,2007FlowerHarris}, 
\begin{equation}
{\rm H_{2} + H} \rightarrow {\rm H + H + H},
\end{equation}
as in the latter case, one is forced to extrapolate the collisional dissociation rate to a temperature range far below that for which reliable experimental values have been measured. Previous studies have shown that the behavior of primordial gas at densities $n > 10^{8} \: {\rm cm^{-3}}$ is highly sensitive to the treatment of three-body H$_{2}$ formation \citep{2011Turkb,2014Bovino} and so it is important to model this process as accurately as possible.

To model H$_{2}$ line cooling in the optically thick limit, we use the Sobolev approximation, as in \citet{2006Yoshida} and \citet{2011bClark}. Although not as accurate as methods involving the solution of the full non-LTE radiative transfer equation \citep{2014Greif} or computation of the H$_{2}$ column density distribution \citep{2015aHartwig}, it has a much lower computational cost, which is an important benefit in our current study given the number of simulations that we run.

In the course of carrying out our simulations, we found that it became extremely computationally expensive to track the non-equilibrium deuterium chemistry in gas with $n \gg 10^{8} \, {\rm cm^{-3}}$, owing to the short chemical timescales involved. Since the HD/H$_{2}$ ratio in this regime just tracks the cosmological ratio of D to H, and since the HD molecules are not important coolants at these densities \citep{2009GloverSavin}, we deal with this problem by switching off explicit tracking of the deuterium chemistry at densities $n > 10^{8} \: {\rm cm^{-3}}$. Instead, we assume that in this regime the ratio of the fractional abundances of HD and H$_{2}$,  $x_{\rm HD} / x_{\rm H_{2}}$, is given by the cosmological D to H ratio, $x_{\rm D, tot} = 2.6 \times 10^{-5}$. We do not expect this computational simplification to have any significant impact on our results.

Finally, in our simulations, we account for the fact that in warm gas with a high molecular fraction, the adiabatic index of the gas is not necessarily equal to the value for a monatomic gas, $\gamma=5/3$, but instead depends on the chemical composition and temperature of the gas. \textsc{arepo} already supports the use of a variable adiabatic index in its HLLD Riemann solver, so our main modification here was to provide routines to compute $\gamma$ as a function of chemical composition and $T$ for a primordial gas. This is carried out along the same lines as in \citet{Boley2007}.

\subsection{Sink particles}
We use sink particles to represent regions of gas collapsing to scales that are smaller than we can feasibly resolve in our current simulations. Sink particles are artificial, non-gaseous particles that are introduced in place of dense concentrations of gas and that can continue to accrete gas from their surroundings, but that otherwise interact only gravitationally. 
They were first introduced for studying present-day star formation by \citet{1995Bate} and have since become a widely-used tool in the computational study of star formation. In the simulations presented here, we use a sink particle module that we have recently developed for {\sc arepo} for use in simulations of both primordial and present-day star formation 
\citep[see also][]{Tress2019,Smith2019}.

In order  to convert a given Voronoi cell into a sink particle, the cell and its surroundings must satisfy a number of conditions. 
First of all, the candidate cell must have a density higher than the threshold density for sink formation, $n_{\rm th}$. Second, it must be at a local minimum of the gravitational potential. Third, the value of the velocity divergence $\nabla \cdot \mathbf{v}$ at the location of the cell must be negative. Fourth, it must not be within the accretion radius $r_{\rm acc}$ of an existing sink; dense gas that is within the accretion radius of a sink is instead a candidate for accretion, as we describe below. From the cells that satisfy all four of these conditions, we next select the densest one and carry out a further series of checks. We compute the total thermal, gravitational and kinetic energy within a control volume of radius $r_{\rm acc}$ surrounding the sink candidate and check that $|E_{\rm grav}| > 2 \, E_{\rm therm}$ (implying that the mass within the control volume exceeds the local Jeans mass) and that the total energy $E_{\rm grav} + E_{\rm kin} + E_{\rm therm} < 0$ (implying that the gas is gravitationally bound). Finally, we also check that $\nabla \cdot \mathbf{v} < 0$ and $\nabla \cdot \mathbf{a} < 0$ within the control volume, where $\mathbf{a}$ is the acceleration. This last check helps to distinguish regions that are truly collapsing from ones that have transiently reached a high density due to tidal interactions or shocks.  If a gas cell fulfils all of these criteria, it is converted to a sink particle with the same mass and momentum. This sink particle is assigned a gravitational softening length equal to one third of the accretion radius in order to avoid artificial fragmentation.

Once created, sink particles can accrete gas from their surroundings. In order for the gas in a Voronoi cell to be eligible for accretion, it must once again satisfy a number of conditions. First, the cell must be located within the accretion radius of sink. Second, the gas density must exceed $n_{\rm th}$. Third, the gas must be gravitationally bound to the sink, i.e.\ its kinetic energy in the rest frame of the sink must be smaller than its gravitational binding energy with the sink. Fourth, it must be moving towards the sink, which is easily verified by computing its radial velocity in the sink rest frame. Finally, its radial acceleration in the sink rest frame must also be negative. If the sinks are highly clustered, with overlapping accretion radii, then cells may sometimes satisfy these conditions with respect to multiple sinks. In that case, we associate the cell with the sink particle to which it is most strongly gravitationally bound. Once we have identified the cells that are candidates for accretion, we next compute how much gas will be accreted by each sink from each cell. We limit the total gas mass that can be accreted from any given cell to either the mass required to reduce the cell density to $n_{\rm th}$ or 90\% of the total gas mass in the cell, whichever is the smaller. The accreted mass and associated momentum are then removed from the gas cell and added to the sink, so that globally mass and momentum are conserved. 

Since {\sc arepo} uses a hierarchical time-stepping scheme, different Voronoi cells can have different timesteps and hence only a subset of cells may be active on any given timestep. Gas is allowed to accrete onto a sink only from active Voronoi cells and only when the sink itself is active. However, the sink timestep is chosen to be the same as the shortest gas cell timestep so that we do not miss accretion from cells that spend only short periods within $r_{\rm acc}$. 

In the simulations presented in this paper, we set $r_{\rm acc} = 2$~AU and $n_{\rm th} = 2.4 \times 10^{15} \: {\rm cm^{-3}}$. The computational cost of simulations such as ours increases significantly as we decrease $r_{\rm acc}$, and a value of 2~AU was the smallest that we found to be practical given the number of simulations that we wanted to run and the time for which we followed fragmentation in each system. 

\subsection{Accretion luminosity}
\label{sec:Lacc}
As gas accretes onto the newly-formed protostars, it releases a significant amount of energy. We account for this by computing the accretion luminosity for each accreting protostar,
\begin{equation}
L_{\mathrm{acc}} = \frac{G M_{\star} \dot{M}}{R_{\star}},
\end{equation}
and the corresponding volumetric heating rate, 
\begin{equation}
\Gamma_{\mathrm{acc}} = \rho \kappa_{\rm P} \left( \frac{L_{\mathrm{acc}}}{4 \pi r^2} \right) \, \rm erg \, s^{-1} \, cm^{-3}.  \label{acc_heat}
\end{equation}
Here, $G$ is the gravitational constant, $\dot{M}$ is the protostellar accretion rate (which we take to be equal to the instantaneous accretion rate onto the sink), $M_{\star}$ and $R_{\star}$ are the mass and radius of the protostar, 
$\rho$ is the gas density,  $\kappa_{\rm P}$ is the Planck mean opacity (see more details below), and $r$ is the distance between the accreting sink and the cell in which we are computing the heating rate. Equation~\ref{acc_heat} assumes that the gas is optically thin to the radiation from the accreting protostar, which is a good approximation in primordial gas \citep{2011Smith}. 

To compute the value of $R_{\star}$, we follow the same approach as in \citet{2011Smith} and adopt the analytical prescriptions given by \citet{2003OmukaiPalla}:
\begin{equation}
\label{eq:radius1}
R_{\star} \propto
\begin{cases}
26 M_{\star}^{0.27} (\dot{M}/10^{-3})^{0.41} \; \; &M_{\star} \leq p_1, \\
A_1 M_{\star}^3 \; \; &p_1 \leq M_{\star} < p_2, \\
A_2 M_{\star}^{-2} \; \; &p_2 \leq M_{\star} \, \& \, R < R_{\rm ms},
\end{cases}
\end{equation}
where $R_{\star}$, $M_{\star}$ and $\dot{M}$ are expressed in units of R$_{\odot}$,
M$_{\odot}$ and ${\rm M_{\odot} \, yr^{-1}}$, respectively, and where the transition points 
$p_{1}$ and $p_{2}$ are given by
\begin{align}
\label{eq:radius2}
p_1 &= 5 (\dot{M}/10^{-3})^{0.27} \, \, \rm M_{\odot}, \\
p_2 &= 7 (\dot{M}/10^{-3})^{0.27} \, \, \rm M_{\odot}. 
\end{align}
Finally, the main sequence radius $R_{\rm ms}$ is given by 
\begin{equation}
\label{eq:radius3}
R_{\mathrm{ms}} = 0.28 M_{\star}^{0.61} \rm R_{\odot}.
\end{equation}

Our approach assumes a quasi-spherical accretion flow onto the sink. We cannot resolve the protostellar surface and the immediate region around our sink due to our choice of the sink accretion radius $r_{\mathrm{acc}} = 2 \, \rm AU$. Therefore, we do not know whether a thin inner disk forms and how thermally efficient the flow from the disk onto the protostellar surface is \citep{2010Hosokawa}. In our studies, however, we generally find thick disks down to our assumed accretion radius \citep[see also][for similar findings]{2011aClark}, in which the sink is embedded, and thus our approximation is reasonable. 

The other assumption made here is that the radius of the protostar responds instantly to changes in the accretion rate. In reality, the characteristic time-scale of the protostellar radial evolution is of the order of $10^3 \, \rm yr$. For the early protostellar evolution dominated by mass growth ($M \lesssim 10 \, \rm M_\odot$), this is governed by the accretion time-scale, while later stages follow the Kelvin-Helmholtz contraction time-scale, that is the time over which the star can internally redistribute the entropy which it has taken up through accretion. However, work by \citet{2011Smith,2012Smith}  indicates that a highly variable accretion rate does not introduce an enormous error in the computation of the radial evolution of the protostar. Moreover, strong variations in the computation of the accretion luminosity due to a time-varying accretion rate are partially compensated for by changes in the protostellar radius (which increases with increasing $\dot{M}$) since $L_{\mathrm{acc}} \propto \dot{M} / R_{\star}$. 

To compute the Planck mean opacity, we again follow \citet{2011Smith} and make use of the values tabulated as a function of density and temperature by \citet{2005MayerDuschl}. One simplification made in that work is the assumption that the gas and radiation temperatures were equal. This does not lead to a large error for gas temperatures of a few thousand degrees, as these are comparable to the actual photospheric temperatures of the accreting protostars. However, it leads to an unrealistically high value for $\kappa_{\rm P}$ and $\Gamma_{\rm acc}$ in gas with $T \sim 10^{4} \: {\rm K}$ or higher. To avoid this problem, we compute $\kappa_{\rm P}$ using a temperature given by
\begin{equation}
T_{\kappa} = {\rm max} \left(T, 6000 \, {\rm K} \right),
\end{equation}
where $T$ is the actual gas temperature.

\section{Initial conditions}
\label{Sec:IC}
The collapse simulations are performed in a box with sidelength $13 \, \rm pc$. We start with $128^3$ cells and let the code evolve with Jeans (de-)refinement making sure that the Jeans length is always resolved by at least 16 cells. We chose to use this number as test simulations carried out using higher numbers of cells per Jeans length proved to be too costly in computational terms to carry on for long beyond the formation of the first sink particle. We note that although studies of the initial collapse of the gas have been carried out using higher numbers of cells per Jeans length \citep[see e.g.][]{2012Turk,bovino13,2014Greif}, most previous simulations that followed the subsequent fragmentation of the disk were carried out with mass resolutions that were comparable to or lower than those achieved here.

We use the density profile of a Bonnor-Ebert (BE) sphere \citep{1955Ebert, 1956Bonnor} as the initial condition for our cloud. This density profile is similar to the density profile of the cold, dense cores formed at the center of the  gravitationally collapsing gas in cosmological simulations of Pop.\ III star formation \citep{2002Abel, 2002Bromm, 2008Yoshida}. It follows
\begin{equation}
\rho(r) = 
	\begin{cases}
		f \times \rho_{\mathrm{BE}}(r) & \text{for } r < R_{\mathrm{BE}} \\
		f \times \rho_{\mathrm{BE}}(R_{\mathrm{BE}})  & \text{for } r \geq R_{\mathrm{BE}} 
	\end{cases}
\end{equation}
where $\rho_{\mathrm{BE}}(r)$ is the density distribution of a critical BE sphere (i.e.\ one on the boundary between stability and instability). The parameter $f$ denotes a density enhancement factor which we use here in order to promote the collapse of our sphere \citep{Shu1977}. We use $f = 1.83$ which gives a central particle number density of $n_c = 1.83 \times 10^{4} \, \rm cm^{-3}$, and a central mass density of $\rho_c = f \times \rho_{\mathrm{BE}}(0) = 3.7 \times 10^{-20} \, \rm g \, cm^{-3}$. We set the radius of the sphere to be $R_{\mathrm{BE}} = 1.87 \, \rm pc$, corresponding to a BE sphere with nondimensional radius of $\xi_{\mathrm{BE}} = 6.5$, which defines the critical value for stability. The mass within this radius is then  $M_{\mathrm{BE}} = 2671 \, \rm M_{\odot}$. The temperature inside the sphere is initially $T = 200 \, \rm K$ which gives an adiabatic sound speed of $c_{\rm s} = 1.93 \, {\rm km \, s}^{-1}$ with an initial adiabatic index of $\gamma = 5/3$ (isothermal sound speed: $c_{\rm s, iso} = 1.16 \, {\rm km \, s}^{-1}$). The sphere is located within an environment of uniformly dense gas with $n_{\mathrm{ext}} = n_{\mathrm{BE}}(R_{\mathrm{BE}}) = 1.31 \times 10^{3} \, \rm cm^{-3}$ ($\rho_{\mathrm{ext}} = 2.7 \times 10^{-21} \, \rm g \, cm^{-3}$) and $T_{\mathrm{ext}} = T_{\mathrm{BE}} = 200 \, \rm K$.\\

We run simulations with nine different setups for the BE sphere, differing in the amount of rotational and turbulent kinetic energy present initially in the gas. Our first set of runs have zero rotation and no turbulence -- we term these {\it pure infall} models. We also carry out simulations with rotation but no turbulence, turbulence with no rotation, and a set of mixed models with four different combinations of rotation and turbulence.  An overview of the models is given in Table~\ref{tab:turb_config}. 

We include rotation by letting our initial cloud rotate around the $z$-axis at a uniform angular velocity $\Omega_0$. By changing $\Omega_{0}$, we can adjust the ratio of rotational kinetic energy to gravitational potential energy for the cloud, $\beta_{\mathrm{rot}} = E_{\mathrm{rot}} / |E_{\mathrm{grav}}|$. We choose $\beta=0.01$ and $\beta=0.1$ as guide values to span a reasonable range from small to high levels of rotation. Due to the lack of direct observations of primordial star-forming clouds, our choice of values is on the one side motivated from observations of nearby molecular cloud cores, where one finds typical values of $\beta \sim 0.02$ \citep{1993Goodman, 2002Caselli}, and on the other side by results from  cosmological simulations of Pop.\ III star formation, which find $\beta \lesssim 0.1$ \citep{2002Bromm, 2006Yoshida}. Our choice also agrees with the values found in the study of \citet{2014Hirano}.

In runs with turbulence, we add a homogeneous, isotropic Gaussian random velocity component onto our BE sphere. We do not drive the turbulence but let it freely decay over the course of the simulation. The Gaussian random field is constructed by a combination of randomly distributed phases together with an amplitude following a Rayleigh distribution \citep{1986Bardeen}. The power spectrum of the velocity field is chosen to be $P(k) \propto k^{-4}$, so that most of the power is contained in large-scale modes. We construct the field in Fourier space, transform it into real space and only use its real parts. We use a mixed mode spectrum, i.e. a spectrum consisting of both solenoidal and compressive modes
\citep[see e.g.][]{Federrath2010}. After deriving the root-mean square value of this field, we invert this value and use it to rescale the velocity of our turbulent field as desired. We carry out runs with two different values of the turbulent $\alpha$ parameter, $\alpha_{\mathrm{turb}} = E_{\mathrm{turb}} / |E_{\mathrm{grav}}|$: a low turbulence case with $\alpha_{\rm turb}=0.05$ and a high turbulence case with $\alpha_{\rm turb}=0.25$. Our choice of the values here is motivated by the studies from \citet{2004aGoodwin,2004bGoodwin}. 

For our chosen initial temperature of 200~K, the ratio between thermal and gravitational energy of our sphere, $\alpha_{\mathrm{therm}} = E_{\mathrm{therm}} / |E_{\mathrm{grav}}| \simeq 0.451$.\footnote{One of the reasons to increase the density by some factor $f$ was to make sure that our sphere would still collapse when we introduced strong rotation ($\beta_{\mathrm{rot}} = 0.1$) and turbulence ($\alpha_{\mathrm{turb}} = 0.25$), i.e.\ to ensure that the total energy of the system is still negative, $E_{\mathrm{tot}} =  E_{\mathrm{grav}} + E_{\mathrm{therm}} + E_{\mathrm{rot}} + E_{\mathrm{turb}}< 0$.}
One can estimate the number of Jeans masses within the sphere by $(E_{\mathrm{therm}} / |E_{\mathrm{grav}}|)^{-3/2}$ which gives us here roughly 3 Jeans masses. 

We carry out 5 realizations per setup. For setups including turbulence each realization has its own random seed. We model the 5 different realizations for setups without turbulence by first choosing 5 uniformly dense boxes which only differ in their random mesh cell positions and then initializing our BE density distribution on them. The naming convention we use for the different setups is of the form $\alpha{M}\beta{N}$, where $M$ indicates the value of $\alpha_{\rm turb}$, with $M = 005$ or 025 corresponding to runs with $\alpha_{\rm turb} = 0.05$ and 0.25, respectively, and $N$ indicates the value of $\beta_{\rm rot}$ using a similar convention. The value of $\alpha$ or $\beta$ is omitted if it is zero. The individual realizations are denoted by a numerical suffix: e.g.\ $\alpha025\beta01$ -- 3 is realization number 3 of setup $\alpha025\beta01$.

We assume a redshift of $z=20$ for our simulations. However, as the gas in our simulations is always much warmer than the cosmic microwave background and the effects of Compton cooling are negligible, the results we obtain should be largely independent of redshift. For our initial elemental abundances we use $x_{\rm He} = 0.079$ and $x_{\rm D} = 2.6 \times 10^{-5}$ \citep{2008GloverAbel}. The values of the fractional abundances of $\mathrm{H_2}$, $\mathrm{H^+}$, $\mathrm{HD}$ and $\mathrm{D^+}$ are motivated by cosmological simulations of Population III.1 star formation, i.e.\ Pop.\ III star formation occurring in minihalos that have not been affected by stellar feedback \citep[e.g.][]{2008Greif}: $x_{\mathrm{H_2}}=1.0 \times 10^{-3}$, $x_{\mathrm{H^+}}=1.0 \times 10^{-7}$, $x_{\mathrm{HD}}=3.0 \times 10^{-7}$ and $x_{\mathrm{D^+}}=2.6 \times 10^{-12}$. All hydrogen and deuterium not included in one of these forms is assumed to be present as neutral atomic hydrogen or deuterium, as appropriate, while all of the helium is assumed to be neutral and atomic. We also assume that the gas is in charge balance, with a free electron abundance given by $x_{\rm e^{-}} = x_{\rm H^{+}} + x_{\rm D^{+}}$. The values of the H$^{-}$ and H$_{2}^{+}$ abundances are not specified in our initial conditions, as these species are assumed to be in chemical equilibrium and so their equilibrium abundances are computed on the fly within the chemistry module. We summarize our initial conditions in Table \ref{tab:IC}. 

\begin{table}
\footnotesize
\centering
\caption{Overview of simulation setups.}
\label{tab:turb_config}
\begin{tabular}{@{}lccccccc@{}}
\toprule
\toprule
Setup                & $\alpha_{\mathrm{turb}}$   &   $\beta_{\mathrm{rot}}$  & $v_{\mathrm{rms}}$ & $\Omega_0$  \\ 
                            &   &    &  $\rm [km \, s^{-1}]$ & $\rm [s^{-1}]$      \\ \midrule
pure infall                    & -       &  -  & - & - \\
$\beta01$                    & -       &  0.1   & - & $3.1 \times 10^{-14}$\\
$\beta001$                  & -       &  0.01 & - & $9.8 \times 10^{-15}$\\
$\alpha025$                & 0.25  &  -      & 1.1 & - \\
$\alpha005$                & 0.05  &  -      & 0.5 & - \\
$\alpha025\beta01$    & 0.25  &  0.1   & 1.1 & $3.1 \times 10^{-14}$ \\
$\alpha025\beta001$  & 0.25  &  0.01 & 1.1& $9.8 \times 10^{-15}$\\ 
$\alpha005\beta01$    & 0.05  &  0.1   & 0.5 & $3.1 \times 10^{-14}$\\ 
$\alpha005\beta001$  & 0.05  &  0.01 & 0.5 & $9.8 \times 10^{-15}$\\ 
\bottomrule
\end{tabular}
\end{table}

\begin{table}
\centering
\caption{Summary of the initial conditions for our simulations.} 
\label{tab:IC}
\begin{tabular}{@{}ll@{}}
\toprule
\toprule
Parameter               & Value   \\  \midrule
$\rho_c$ & $3.7 \times 10^{-20} \, \rm g \, cm^{-3}$  \\
$\rm M_{BE}$ & $2671 \, \rm M_{\odot}$ \\
$\rm R_{BE}$ & $1.87 \, \rm pc$\\
$\rm T_{in}$ & 200 K\\
$x_{\mathrm{H_2}}$ & $1.0 \times 10^{-3}$ \\ 
$x_{\mathrm{H^+}}$ & $1.0 \times 10^{-7}$ \\
$x_{\mathrm{HD}}$ & $3.0 \times 10^{-7}$ \\
$x_{\mathrm{D^+}}$ & $2.6 \times 10^{-12}$ \\
\bottomrule
\end{tabular}
\end{table}

\section{Results}
\label{Sec:Results}

\subsection{Formation time of the first sink}
\begin{figure}
\centering 
\includegraphics[width=0.45\textwidth]{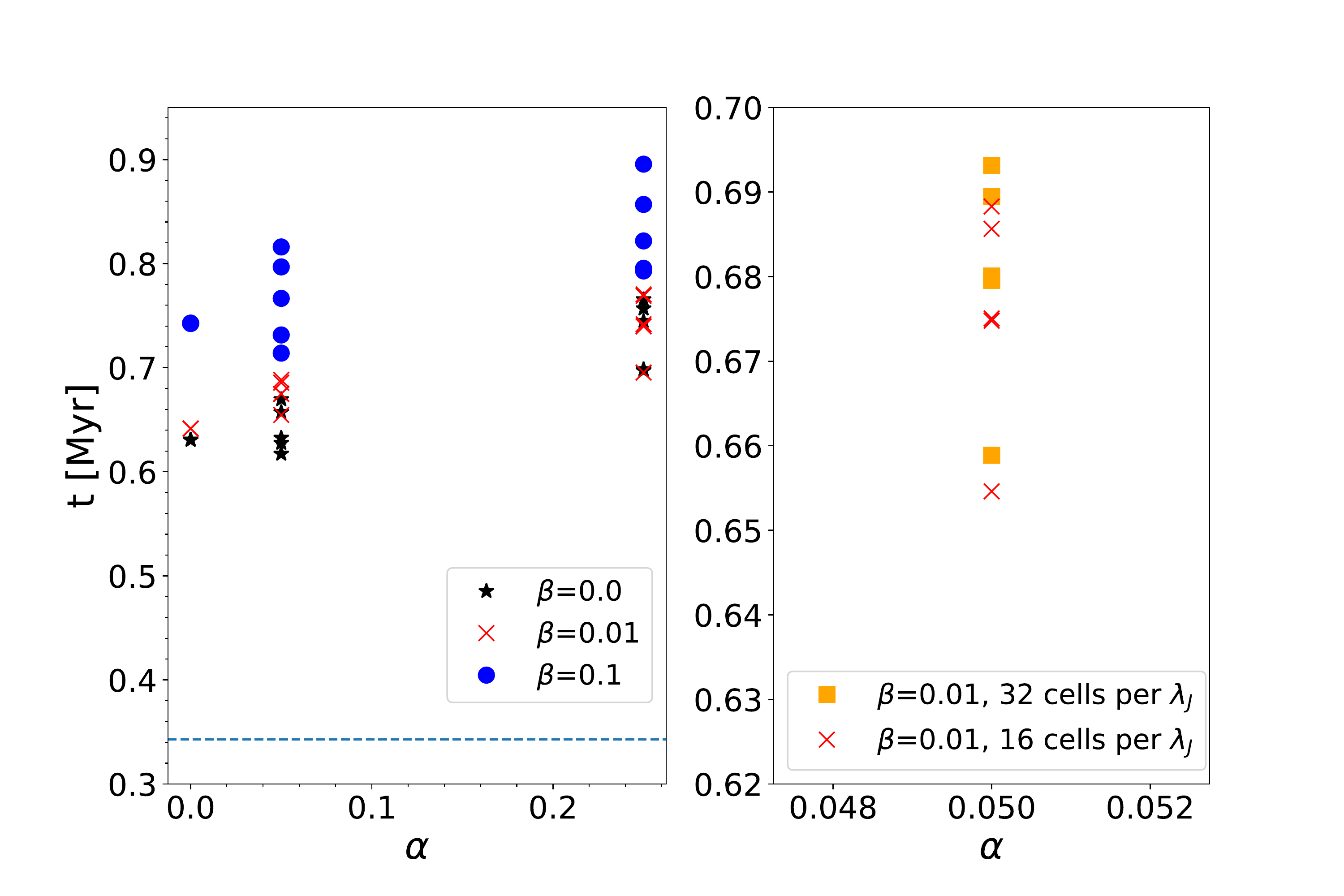} 
\caption{Formation time of the first sink particle in each simulation, plotted as a function of $\alpha_{\rm turb}$. The left-hand panel shows the results for several different values of the rotational $\beta$ parameter.  The right-hand panel shows a comparison of the results of setup \textit{$\alpha005\beta001$} for simulations using our standard Jeans refinement criterion, for which the Jeans length is always resolved by at least 16 cells (\textit{red crosses}) and simulations using a more stringent refinement criterion of 32 cells per Jeans length (\textit{orange squares}).  The blue dashed line in the left-hand panel indicates the free-fall time of the cloud at its initial central density.
\label{fig:creation_time}} 
\end{figure}

We start our survey of the results of our large set of simulations by looking at the simple question of how long it takes to form the first sink particle in each of our simulations. The results are listed in Table~\ref{tab:turb_result} and summarized in Figure~\ref{fig:creation_time}. We see immediately that both a higher initial level of rotation and a higher initial level of turbulence within the collapsing gas cloud delays the onset of star formation. Similar trends were observed in previous studies \citep[see e.g][]{2011bClark, 2018Riaz}. 

In the runs without any turbulence ($\alpha=0$), the formation time of the first sink is almost identical in each realization of a given setup, with the largest difference being $\sim 200 \, \rm yr$ between the formation times of the first sinks in runs $\beta01$-1 and $\beta01$-2. This suggests that the small amount of numerical noise introduced by the variations in the initial cell distribution in the zero turbulence runs does not have a significant effect on the collapse timescale. On the other hand, the runs with turbulence show a significant spread in the sink formation times, amounting to $\sim 0.1$~Myr in most cases. This is also clear in Table \ref{tab:turb_result}, where we list the time at which the first sink forms ($t_{\rm SF}$) for each run. The dashed blue line in Fig. \ref{fig:creation_time} shows the free-fall time of the cloud, $t_{\mathrm{ff}} = \sqrt{3 \, \pi / (32 \, G \, \rho_{\mathrm{c}})} \simeq 0.34 \, {\rm Myr}$, i.e.\ the time that it would take for a spherical cloud with a density equal to the initial central density of our Bonnor-Ebert sphere to collapse to a point in the absence of thermal pressure. We see that even in the  {\it pure infall} runs, the time that elapses before the first sink forms is $\sim 2 t_{\rm ff}$, demonstrating that pressure forces play a significant role in slowing the collapse of the gas. Including additional support in the form of non-zero turbulence or rotation further slows the collapse, but overall is less important than the support provided by the thermal pressure. 
 
We have also examined whether our choice of resolution has any significant effect on the collapse time. \citet{2012Turk} argue that a resolution of 16 cells per Jeans length -- our fiducial value -- is insufficient for resolving the small-scale gravitational collapse of primordial star-forming gas, and that at least 32-64 cells per Jeans length is required. However, \citet{bovino13} show that this result is largely due to the use of a first order ordinary differential equation (ODE) solver for the coupled cooling and chemistry ODEs in the \citet{2012Turk} study, and that simulations carried out using a higher-order ODE solver display far less dependence on numerical resolution. To investigate this, we resimulated the five different realizations of the $\alpha005\beta001$ setup using 32 cells per Jeans length. The results are shown in the right-hand panel in Figure~\ref{fig:creation_time}. Consistent with \citet{bovino13}, we see that increasing the resolution from 16 to 32 cells per Jeans length has only a minor effect on the collapse. The collapse time lengthens slightly in the higher resolution run, but only by $\sim 0.01$~Myr, which is less than the scatter between the different random realizations of the initial conditions.

\citet{2012Turk} also note that very high resolution is required in their runs in order to follow the amplification of the magnetic field by the turbulent dynamo \citep[see also][]{2011bFederrath}. However, this requirement is not relevant for our simulations, which do not include a magnetic field.

\begin{table*}
\footnotesize
\centering
\caption{Overview of the detailed results of all our simulations regarding the time at which the first sink forms, $t_{\mathrm{SF}}$, the final year considered for our analysis, $t_{\mathrm{final}}$, the total number of sink particles formed by that time, $N_{\mathrm{tot}}$, and the corresponding total mass in sinks, $M_{\mathrm{tot}}$. For our analysis of the sink mass functions, we also list $t_{\mathrm{same}}$, the time at which the total mass of all sinks per simulations reaches $\sim 36 \, \rm M_{\odot}$, and $M_{\mathrm{same}}$, the exact total mass in sinks at that time.}
\label{tab:turb_result}
\begin{tabular}{llcclcccc}
\hline
Realization           & $\alpha_{\mathrm{turb}}$ & $\beta_{\mathrm{rot}}$ & $t_{\mathrm{SF}}$/Myr & $t_{\mathrm{final}}$/yr & $t_{\mathrm{same}}$/yr & $N_{\mathrm{tot}}$ & $M_{\mathrm{tot}}/M_{\odot}$ & $M_{\mathrm{same}}/M_{\odot}$ \\ \hline
pure infall-1         & -                        & -                      & 0.631                 & $989$                   & $448$                  & $1$                & $69.9$                       & $36.6$                        \\
pure infall-2         & -                        & -                      & 0.631                 & $948$                   & $439$                  & $1$                & $67.8$                       & $36.0$                        \\
pure infall-3         & -                        & -                      & 0.631                 & $699$                   & $444$                  & $1$                & $53.2$                       & $36.4$                        \\
pure infall-4         & -                        & -                      & 0.631                 & $443$                   & $443$                  & $1$                & $36.3$                       & $36.3$                        \\
pure infall-5         & -                        & -                      & 0.631                 & $735$                   & $449$                  & $1$                & $55.4$                       & $36.7$                        \\ \hline
$\beta01-1$           & -                        & 0.1                    & 0.742                 & $1002$                  & $970$                  & $49$               & $37.2$                       & $36.5$                        \\
$\beta01-2$           & -                        & 0.1                    & 0.743                 & $1001$                  & $935$                  & $31$               & $38.3$                       & $36.4$                        \\
$\beta01-3$           & -                        & 0.1                    & 0.743                 & $1003$                  & $1151$                 & $42$               & $33.0$                       & $36.4$                        \\
$\beta01-4$           & -                        & 0.1                    & 0.743                 & $1003$                  & $1003$                 & $27$               & $36.4$                       & $36.4$                        \\
$\beta01-5$           & -                        & 0.1                    & 0.743                 & $1005$                  & $1005$                 & $45$               & $36.3$                       & $36.3$                        \\ \hline
$\beta001-1$          & -                        & 0.01                   & 0.641                 & $1003$                  & $608$                  & $28$               & $58.9$                       & $36.5$                        \\
$\beta001-2$          & -                        & 0.01                   & 0.642                 & $1003$                  & $595$                  & $26$               & $51.6$                       & $36.2$                        \\
$\beta001-3$          & -                        & 0.01                   & 0.641                 & $1003$                  & $591$                  & $37$               & $59.4$                       & $36.2$                        \\
$\beta001-4$          & -                        & 0.01                   & 0.642                 & $1003$                  & $610$                  & $43$               & $55.9$                       & $36.2$                        \\
$\beta001-5$          & -                        & 0.01                   & 0.641                 & $1005$                  & $602$                  & $35$               & $57.5$                       & $36.8$                        \\ \hline
$\alpha025-1$         & 0.25                     & -                      & 0.698                 & $995$                   & $325$                  & $72$               & $67.9$                       & $36.4$                        \\
$\alpha025-2$         & 0.25                     & -                      & 0.757                 & $992$                   & $578$                  & $40$               & $54.8$                       & $36.5$                        \\
$\alpha025-3$         & 0.25                     & -                      & 0.766                 & $1008$                  & $511$                  & $45$               & $64.6$                       & $36.1$                        \\
$\alpha025-4$         & 0.25                     & -                      & 0.697                 & $1011$                  & $642$                  & $19$               & $52.1$                       & $36.5$                        \\
$\alpha025-5$         & 0.25                     & -                      & 0.745                 & $1008$                  & $352$                  & $118$              & $75.1$                       & $36.4$                        \\ \hline
$\alpha005-1$         & 0.05                     & -                      & 0.670                 & $1003$                  & $369$                  & $61$               & $87.7$                       & $36.1$                        \\
$\alpha005-2$         & 0.05                     & -                      & 0.617                 & $1002$                  & $569$                  & $57$               & $68.2$                       & $36.7$                        \\
$\alpha005-3$         & 0.05                     & -                      & 0.633                 & $1005$                  & $599$                  & $63$               & $60.7$                       & $36.3$                        \\
$\alpha005-4$         & 0.05                     & -                      & 0.657                 & $1002$                  & $355$                  & $51$               & $74.1$                       & $36.1$                        \\
$\alpha005-5$         & 0.05                     & -                      & 0.628                 & $1008$                  & $502$                  & $39$               & $61.7$                       & $36.1$                        \\ \hline
$\alpha025\beta01-1$  & 0.25                     & 0.1                    & 0.796                 & $1007$                  & $479$                  & $41$               & $53.4$                       & $36.4$                        \\
$\alpha025\beta01-2$  & 0.25                     & 0.1                    & 0.857                 & $1012$                  & $1585$                 & $32$               & $25.2$                       & $36.3$                        \\
$\alpha025\beta01-3$  & 0.25                     & 0.1                    & 0.822                 & $1008$                  & $1023$                 & $21$               & $34.8$                       & $36.0$                        \\
$\alpha025\beta01-4$  & 0.25                     & 0.1                    & 0.793                 & $1000$                  & $575$                  & $53$               & $54.4$                       & $36.1$                        \\
$\alpha025\beta01-5$  & 0.25                     & 0.1                    & 0.900                 & $1015$                  & $738$                  & $41$               & $50.7$                       & $36.7$                        \\ \hline
$\alpha025\beta001-1$ & 0.25                     & 0.01                   & 0.769                 & $1000$                  & $753$                  & $34$               & $50.4$                       & $36.2$                        \\
$\alpha025\beta001-2$ & 0.25                     & 0.01                   & 0.695                 & $1000$                  & $369$                  & $55$               & $76.8$                       & $36.9$                        \\
$\alpha025\beta001-3$ & 0.25                     & 0.01                   & 0.742                 & $1032$                  & $751$                  & $52$               & $66.5$                       & $36.4$                        \\
$\alpha025\beta001-4$ & 0.25                     & 0.01                   & 0.770                 & $1008$                  & $173$                  & $61$               & $84.1$                       & $36.6$                        \\
$\alpha025\beta001-5$ & 0.25                     & 0.01                   & 0.740                 & $1010$                  & $617$                  & $44$               & $64.5$                       & $36.9$                        \\ \hline
$\alpha005\beta01-1$  & 0.05                     & 0.1                    & 0.714                 & $1007$                  & $837$                  & $34$               & $42.0$                       & $36.2$                        \\
$\alpha005\beta01-2$  & 0.05                     & 0.1                    & 0.797                 & $1013$                  & $1155$                 & $30$               & $35.4$                       & $36.4$                        \\
$\alpha005\beta01-3$  & 0.05                     & 0.1                    & 0.731                 & $1004$                  & $924$                  & $39$               & $39.0$                       & $36.5$                        \\
$\alpha005\beta01-4$  & 0.05                     & 0.1                    & 0.816                 & $1009$                  & $1860$                 & $15$               & $23.0$                       & $36.4$                        \\
$\alpha005\beta01-5$  & 0.05                     & 0.1                    & 0.767                 & $1007$                  & $1068$                 & $32$               & $33.7$                       & $36.2$                        \\ \hline
$\alpha005\beta001-1$ & 0.05                     & 0.01                   & 0.688                 & $1010$                  & $644$                  & $40$               & $47.8$                       & $36.9$                        \\
$\alpha005\beta001-2$ & 0.05                     & 0.01                   & 0.655                 & $1006$                  & $576$                  & $53$               & $65.0$                       & $36.5$                        \\
$\alpha005\beta001-3$ & 0.05                     & 0.01                   & 0.675                 & $1013$                  & $514$                  & $45$               & $64.0$                       & $36.6$                        \\
$\alpha005\beta001-4$ & 0.05                     & 0.01                   & 0.675                 & $1005$                  & $325$                  & $81$               & $81.4$                       & $36.0$                        \\
$\alpha005\beta001-5$ & 0.05                     & 0.01                   & 0.686                 & $1007$                  & $651$                  & $32$               & $48.4$                       & $36.3$                        \\ \hline
\end{tabular}
\end{table*}

\subsection{Number and mass of sinks formed}
\label{sec:nummass}

\begin{figure}
\centering 
\includegraphics[width=0.48\textwidth]{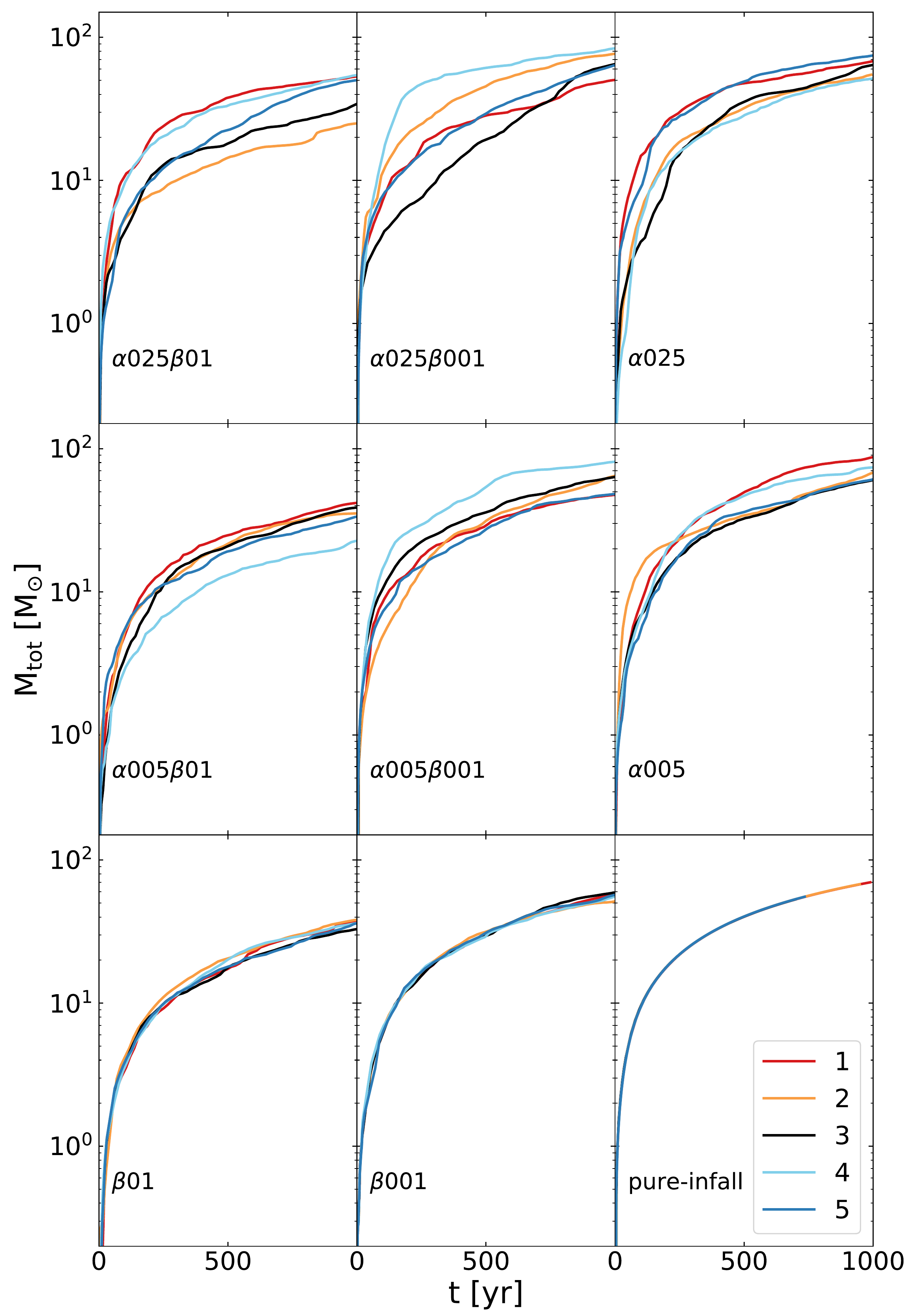} 
\caption{Total mass in sinks as a function of time after the first sink is created. Each panel shows the results for the five different realizations of each setup.
Increasing the initial turbulent or rotational energy tends to systematically decrease the total mass in sinks at any given time. We also see that there is
significant variation from simulation to simulation for all of the setups with non-zero initial turbulence, but not for the setups with non initial turbulence.
\label{fig:mev}} 
\end{figure}

\begin{figure*}
\centering 
\includegraphics[width=0.8\textwidth]{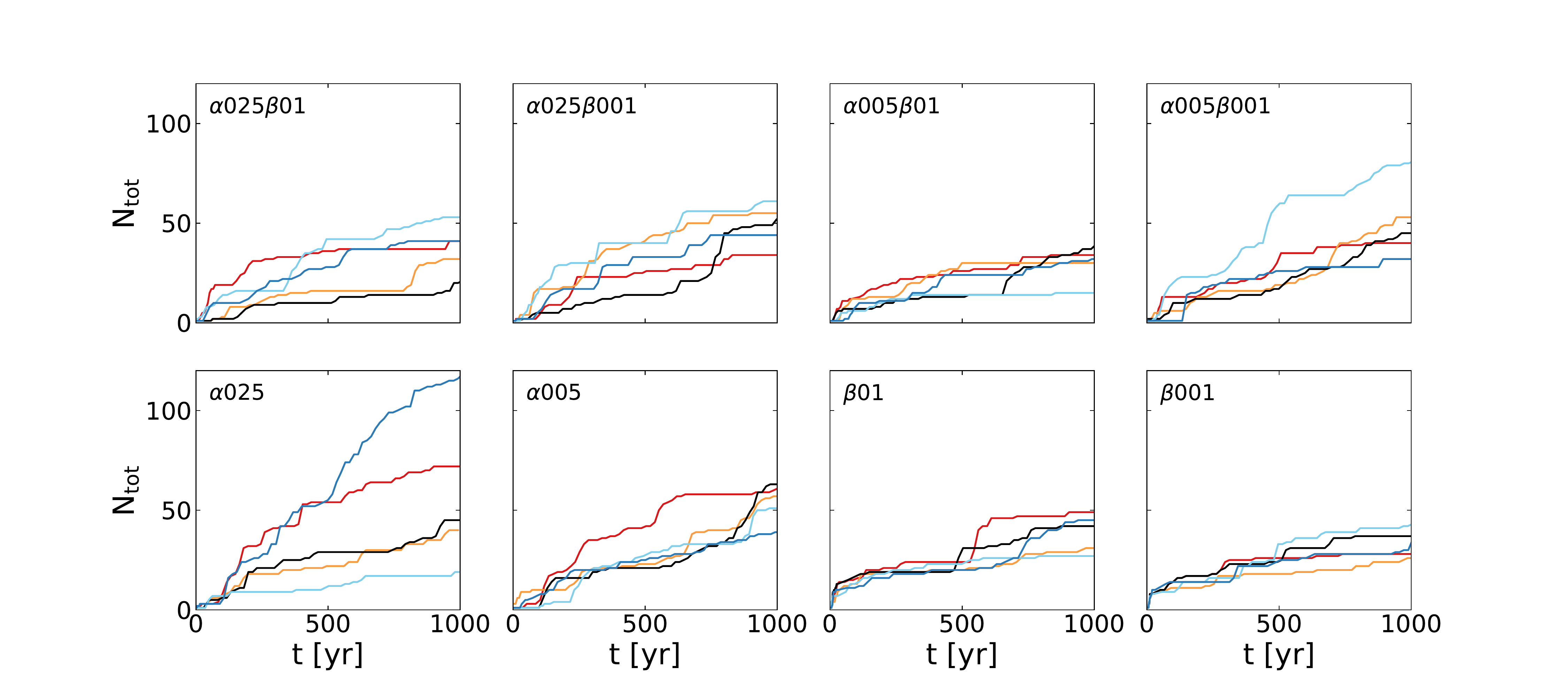} 
\caption{Total number of sinks as a function of the time elapsed after the formation of the first sink. The color scheme is the same as in Fig. \ref{fig:mev}. 
We see that the number of sinks that form varies substantially from simulation to simulation for all of the setups.
\label{fig:Nev}} 
\end{figure*}

We compare most of our runs at a time of $\sim 1000 \, \rm yr$ after first sink formation.\footnote{For technical reasons, the different {\sc arepo} simulations do not produce their final output dumps after precisely 1000~yr, but rather in the interval $t = 1000 \pm 10 \, \rm yr$. In some runs, additional sinks were created shortly after $t = 1000 \, \rm yr$ but before the final output dump. For these runs, we perform our comparisons using data from the previous output dump, produced shortly before $t = 1000$~yr.} At this time, many of the runs contain sinks that are massive enough to start producing ionizing photons once the stars that they represent reach the main sequence. Since our simulations do not currently account for this form of feedback, we stop at this point before the lack of this feedback becomes crucial in changing our results. We stop our {\it pure infall} models earlier than $1000 \, \rm yr$, as these runs require much more computing time than our standard runs. However, this should not greatly affect our analysis, as the behavior of these runs is clear and is unlikely to change significantly in the period between $t_{\rm final}$ and 1000~yr.

\begin{figure*}
\centering 
\includegraphics[width=0.9\textwidth]{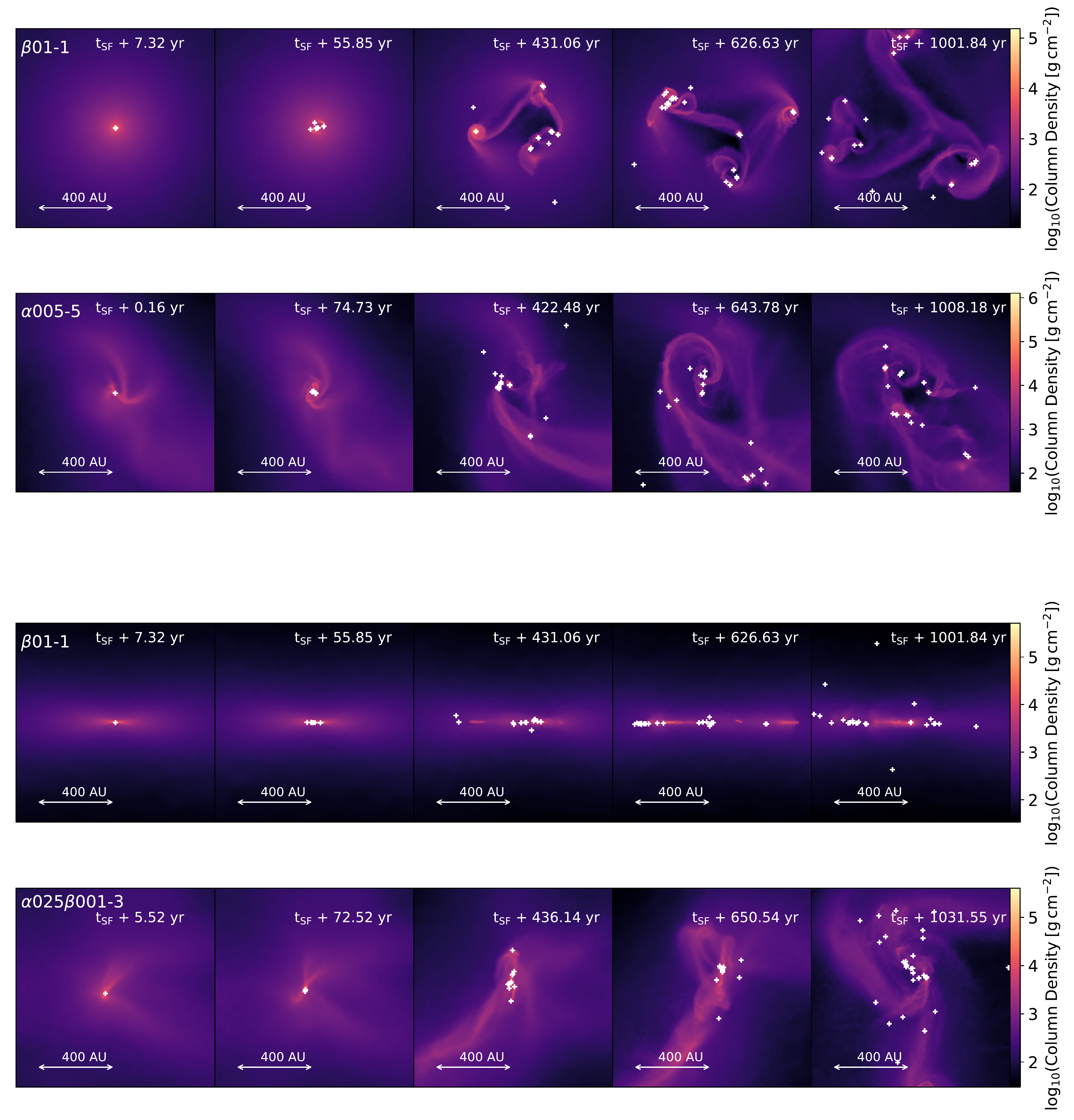} 
\caption{Column density plots (two top rows: $x$-$y$ plane of the simulation box; two bottom rows: $x$-$z$ plane of the simulation box) of the evolution of realization $\beta01$-1, $\alpha005$-5, and $\alpha025\beta001$-3. In the purely rotational runs, the protostellar disk lies in the geometrical $x$-$y$ plane of our simulation box during the whole of the simulation, as demonstrated here for run $\beta01$-1. In all runs including turbulence, however, the orientation of the disk varies strongly over time. The projections are centered on the center of mass of both sink particles and gas within a 500 AU radius around the most massive sink particle at the time of the snapshot considered. The projection thickness is about half the size of our simulation box ($\sim 6.5 \, \rm pc$), the distance between the position of the first sink and the edge of the simulation box. 
\label{fig:projections}} 
\end{figure*}

In the {\it pure infall} realizations only one sink forms and grows rapidly via accretion, reaching a mass of over $30 \, \rm M_{\odot}$ in only $\sim 100$~yr. In contrast, all of the other runs show considerable fragmentation within the first 1000~yr, forming anywhere between 15 and $\sim 120$ sinks. See Fig.~\ref{fig:projections} for illustrative examples of the evolution of some of our realizations. The total number of sinks ($N_{\rm tot}$) and the total mass in sinks ($M_{\rm tot}$) in each realization at the final output time is listed in Table \ref{tab:turb_result}. This difference in behavior occurs because the {\it pure infall} runs are the only ones in which no disk forms over the course of the simulations. Some non-zero angular momentum arises during the gravitational collapse but is too small to trigger the formation of a disk. In all of the other runs\footnote{The initial angular momentum of the gas in the runs with $\alpha > 0$ and $\beta = 0$ is small and varies from realization to realization. In principle, the angular momentum for a particular realization could be zero, but this is vanishingly unlikely.} a dense, gravitationally unstable accretion disk forms which soon fragments.

In Fig. \ref{fig:mev} and \ref{fig:Nev} we plot the evolution of the total mass in sinks, and the total number of sinks respectively. Each panel illustrates one setup, with the five realizations being indicated by different colors. We immediately see in almost all panels  that there is a considerable scatter in the results of each setup. 

In terms of the total mass accreted, we find that the scatter is smallest for the purely rotational and the {\it pure infall} simulations, consistent with the small spread in initial sink formation times in these runs, and largest for the simulations with the highest level of turbulence, particularly in the cases where we also have $\beta > 0$. 

In terms of the total number of sinks formed, we find that there is a significant scatter for every setup (except for the {\it pure infall} case), but that this scatter is clearly larger for the runs with a high level of turbulence. Interestingly, the runs with rotation but no initial turbulence also show clear differences in the number of sinks formed from realization to realization. In this case, the only difference between is the initial positioning of the mesh cells. 
By changing the initial positioning, we change the details of the numerical noise, and this small difference is enough to produce very different fragmentation outcomes, even though the mass going into the fragments varies little from run to run. This result is in good agreement with the study of \citet{Susa2019}, who also found clear differences in the number of fragments formed in simulations of Pop.\ III star formation carried out with the same initial cloud parameters (in their case, rotation with $\beta = 0.05$ but no turbulence) but with different realizations of their initial SPH particle distribution.

This high sensitivity to small changes in the initial conditions suggests that the fragmentation of the disk is a chaotic process, a conclusion which is also consistent with the large amounts of scatter we see in the turbulent runs. If so, then this means that in order to study and derive general trends about the effects of turbulence or rotation on Population III star formation, one needs to consider the statistics of a sample of runs with varying initial conditions (initial mesh cell positioning, random seed used for the turbulence, etc.), rather than just comparing single realizations \citep[see also][who find a similar result in the context of present-day star formation]{2004aGoodwin, 2004bGoodwin}.

\begin{figure*}
\centering 
\includegraphics[width=0.8\textwidth]{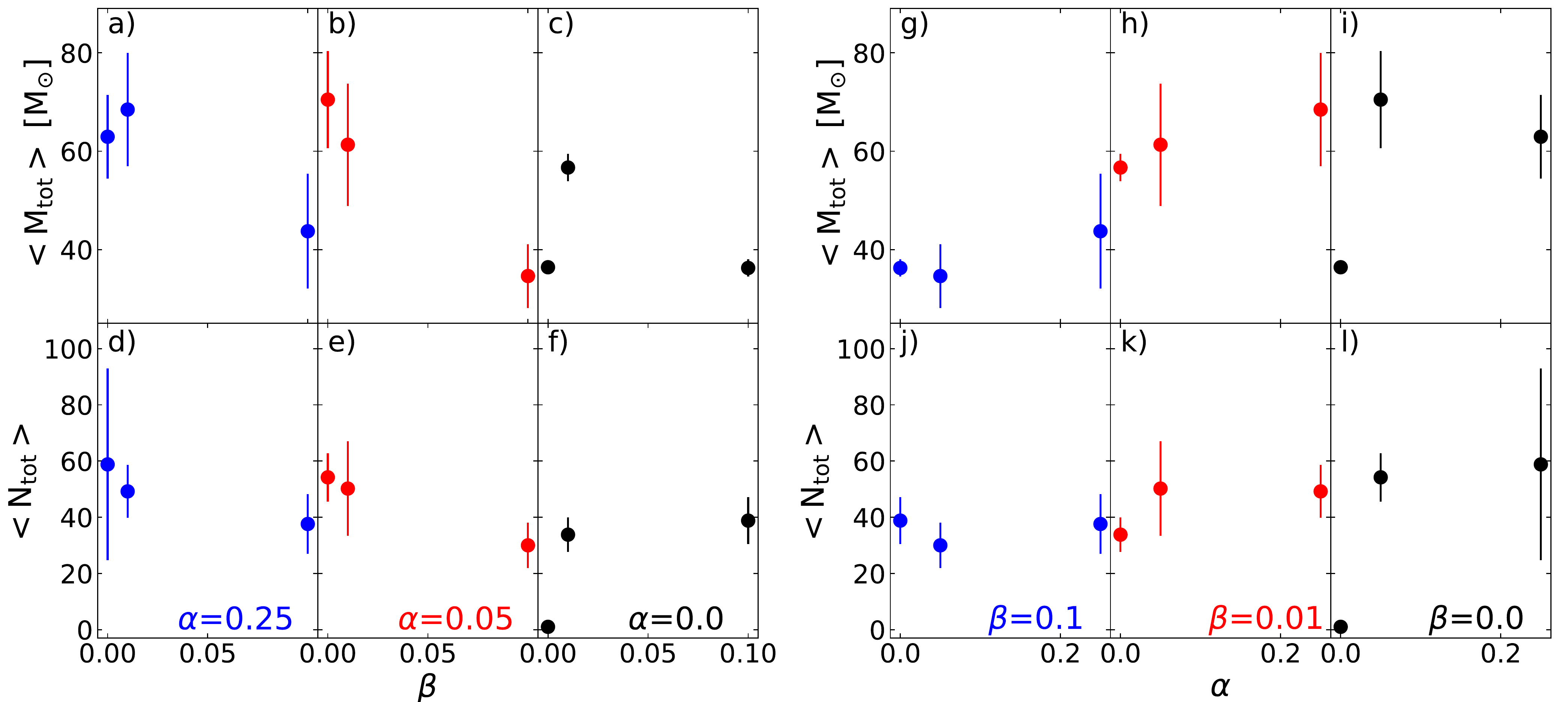} 
\caption{Overview of average final total mass in sinks $\langle M_{\mathrm{tot}} \rangle$ (top panels) and average final total number of sinks $\langle N_{\mathrm{tot}} \rangle$ (bottom panels) per setup versus the rotational $\beta$ (left panels) or the turbulent $\alpha$ (right panels) parameter. Each column corresponds to runs with a fixed value of either $\alpha$ or $\beta$: this is stated in the bottom panel in each column and also indicated by the different colors. Considerable scatter is apparent in the values of $\langle M_{\mathrm{tot}} \rangle$ and $\langle N_{\mathrm{tot}} \rangle$ recovered for a given setup, with the exception of the runs with no initial turbulence, which show little scatter in $\langle M_{\mathrm{tot}} \rangle$  but still moderate scatter in $\langle N_{\mathrm{tot}} \rangle$.
\label{fig:averageplot}} 
\end{figure*}

\begin{table}
\footnotesize
\centering
\caption{Average total number of sinks per setup, $\langle N_{\mathrm{tot}} \rangle$, and average total mass in sinks, $\langle M_{\mathrm{tot}} \rangle$, at $t \sim 1000$~yr, along with the corresponding standard deviations. We note that since not all {\it pure infall} runs were able to run until $t \sim 1000$~yr, we derived $\langle M_{\mathrm{tot}} \rangle$ and $\sigma_{\mathrm{M_{\mathrm{tot}}}}$ of the {\it pure infall} case from the $M_{\mathrm{same}}$ values at the times $t_{\mathrm{same}}$ as given in Table \ref{tab:turb_result}. 
}
\label{tab:average_mass}
\begin{tabular}{lcccc}
\toprule
\toprule
Setup                & $\langle N_{\mathrm{tot}} \rangle$    & $\sigma_{\mathrm{N_{\mathrm{tot}}}}$   &   $\langle M_{\mathrm{tot}} \rangle [\rm M_{\mathrm{\odot}}]$ & $\sigma_{\mathrm{M_{\mathrm{tot}}}}$ \\  \midrule
pure infall 			 & 1      & 0          &    36.4 &   0.2\\ 
$\beta01$ 		 & 38.8 & 8.4      &    36.3 &   1.8\\
$\beta001$ 		 & 33.8 & 6.2    &    56.7 &   2.8\\
$\alpha025$ 		 & 58.8 & 34.1  &    62.9 &   8.5\\
$\alpha005$ 		 & 54.2 & 8.6    &    70.5 &   9.9\\
$\alpha025\beta01$   & 37.6 & 10.7   &   43.7 & 11.7 \\
$\alpha025\beta001$ & 49.2 & 9.4    &    68.5 & 11.5\\ 
$\alpha005\beta01$   & 30.0 & 8.1    &    34.6 &  6.5\\ 
$\alpha005\beta001$ & 50.2 & 16.8  &  61.3 &  12.4\\ 
\bottomrule
\end{tabular}
\end{table}

In Fig. \ref{fig:averageplot}, we plot the average total mass in sinks, $\langle M_{\mathrm{tot}} \rangle$, and the average total number of sinks, $\langle N_{\mathrm{tot}} \rangle$, per setup versus $\beta$ (left subplot) or $\alpha$ (right subplot). The different color-coding of the markers denotes different levels of either constant $\alpha$ or $\beta$ depending on which other one is varied on the $x$-axis. The error bars show the standard deviation in each quantity. We also list the mean and standard deviation in Table \ref{tab:average_mass}. We see directly from this that the scatter in $N_{\mathrm{tot}}$ and $M_{\mathrm{tot}}$ is smaller in the purely rotational runs than in the runs containing initial turbulence. In addition, the mean values of these quantities are also smaller in these runs than in most of the turbulent simulations.

In panel c) and f), the purely rotational data is illustrated while in i) and l) the purely turbulent runs are displayed. We see directly that the purely rotational simulations (together with the {\it pure infall} models) have the smallest error bars of all data points indicating the smallest scatter in results as we have already discussed above. Furthermore, we find that both $\langle N_{\mathrm{tot}} \rangle$ and $\langle M_{\mathrm{tot}} \rangle$ are smaller for the purely rotational models than for the purely turbulent ones. Regarding $\langle M_{\mathrm{tot}} \rangle$ we find a larger value for a lower level of either rotation or turbulence, while for $\langle N_{\mathrm{tot}} \rangle$ the values are similar for both levels.

In the mixed runs, a couple of other trends are obvious. For fixed but non-zero $\alpha$, adding a small amount of initial rotational energy has little effect -- there is no significant difference between $\langle N_{\mathrm{tot}} \rangle$ and $\langle M_{\mathrm{tot}} \rangle$ in the turbulent runs with $\beta = 0$ and in those with $\beta = 0.01$. On the other hand, adding a much larger amount of rotational energy does lead to a clear difference in outcome: both $\langle N_{\mathrm{tot}} \rangle$ and $\langle M_{\mathrm{tot}} \rangle$ are significantly smaller in the runs with $\beta = 0.1$ than in the lower $\beta$ runs, suggesting that in this case the disk is more stable due to a larger differential rotation, i.e. stronger shear motions that allow accumulated gas to be redistributed efficiently thereby keeping the surface density of the disk roughly constant.

Compared to the effect of rotation, a change in the degree of turbulence does not yield very different results in terms of $\langle N_{\mathrm{tot}} \rangle$. 
There is, however, a small  difference between low and high $\alpha$ parameter for $\langle M_{\mathrm{tot}} \rangle$ which can be seen best in panels g) and h). 

\subsubsection{Comparison to previous studies}
\label{sec:average-comparison}
\citet{2011bClark} and \citet{2018Riaz} have also examined the effect of varying the initial turbulent energy in simulations of the collapse of metal-free gas clouds. Since these studies were carried out with different hydrodynamical codes from our own study\footnote{\citet{2011bClark} used the {\sc gagdet} code of \citet{2005Springel}, while \citet{2018Riaz} used the {\sc gradsph} code of \citet{2009Vanaverbeke}.}, it is interesting to compare their results with ours. 

\citet{2011bClark} consider two different scenarios: the collapse of a Bonnor-Ebert sphere with mass $1000 \: {\rm M_{\odot}}$ and initial temperature 300~K, taken to be representative of Pop.\ III.1 star formation, and a Bonnor-Ebert sphere with mass $150 \; {\rm M_{\odot}}$ and initial temperature 75~K, taken to represent Pop.\ III.2 star formation. The first of these setups is clearly much closer to our own and is the one with which we will compare our results. \citet{2018Riaz} consider a single setup, a uniform sphere with mass $1.3041 \times 10^{4} \: {\rm M_{\odot}}$ and initial temperature 300~K.  Both studies quantified the initial turbulent energy of the gas in terms of the ratio between the rms turbulent velocity $v_{\rm rms}$ and the sound speed $c_{\rm s}$. \citet{2011bClark} studied cases with $v_{\mathrm{rms}}/c_{\mathrm{s}} = 0.1, \, 0.2, \, 0.4$ and 0.8, while \citet{2018Riaz} studied cases with $v_{\mathrm{rms}}/c_{\mathrm{s}} = 0.5, 1.0$ and 2.0. In terms of the $\alpha_{\rm turb}$ parameter used in our study, these equate to $\alpha_{\rm turb} \sim 2 \times 10^{-3}, \, 6 \times 10^{-3}, \, 0.03,$ and 0.1 for the \citet{2011bClark} study and $\alpha_{\rm turb} \sim 0.02, \, 0.07,$ and 0.28 for the \citet{2018Riaz} study.

Direct comparison of the numbers and masses of the sinks formed in these studies with the values from our own studies is not informative, owing to the large differences in mass resolution and sink accretion radius: $M_{\rm res} = 0.05 \, {\rm M_{\odot}}$, $r_{\rm acc} = 20$~AU for the \citealt{2011bClark} study and $M_{\rm res} = 1.133 \, {\rm M_{\odot}}$, $r_{\rm acc} = 26$~AU for the \citet{2018Riaz} study, versus an effective mass resolution $M_{\rm res} < 0.01  \, {\rm M_{\odot}}$ and $r_{\rm acc} = 2$~AU in the simulations presented here. However, it is still useful to compare overall trends. 

\citet{2011bClark} find that in their lowest $\alpha_{\rm turb}$ run, the gas does not fragment and the final outcome is very similar to our {\it pure infall} runs. On the other hand, in their other runs, even the relatively small amount of turbulence is enough to lead to significant fragmentation of the gas. Specifically, in these runs the non-zero angular momentum associated with the initial turbulence leads to the formation of a dense, disk-like structure which readily fragments. The relation of the number of fragments formed to the initial turbulent energy is unclear, consistent with our finding that there is considerable scatter from run to run, which can easily overwhelm any weak underlying trend.
 
On the other hand,  \citet{2018Riaz} find that the gas in their runs with $\alpha_{\rm turb} \sim 0.02$ and 0.07 does not fragment, with fragmentation only occurring for their highest $\alpha_{\rm turb}$ run. This stands in stark contrast to our results and those of \citet{2011bClark}. Interestingly, \citet{2018Riaz} report that disk formation does not occur in their low $\alpha_{\rm turb}$ runs, suggesting that the primary reason for the lack of fragmentation in these runs is that their purely turbulent initial conditions start with a much lower level of angular momentum than those in our study or in \citet{2011bClark}. This may be related to the different way in which the turbulent velocity fields are initialized: 
in \citet{2011bClark} and in our simulations, the turbulent velocity field contains a mix of compressive and solenoidal modes and has most of the power concentrated on large scales, while in \citet{2018Riaz} the initial turbulent velocity field is purely solenoidal and has power distributed over a wider range of scales. Further investigation of this point would prove worthwhile, but is out of the scope of our present study.

\citet{2018Riaz} also consider runs with both turbulence and rotation, similar to our mixed runs. In this case, they find that disk formation and fragmentation occurs even when $\alpha_{\rm turb}$ is small. Their results suggest that increasing $\beta$ leads to a more stable disk and hence less fragmentation, in agreement with our findings, although as they only consider one realization of each of their setups, it is impossible to assess the statistical significance of their results. 

\subsection{Mass evolution and accretion behavior}

\begin{figure*}
\centering 
\includegraphics[height=1.2\textwidth]{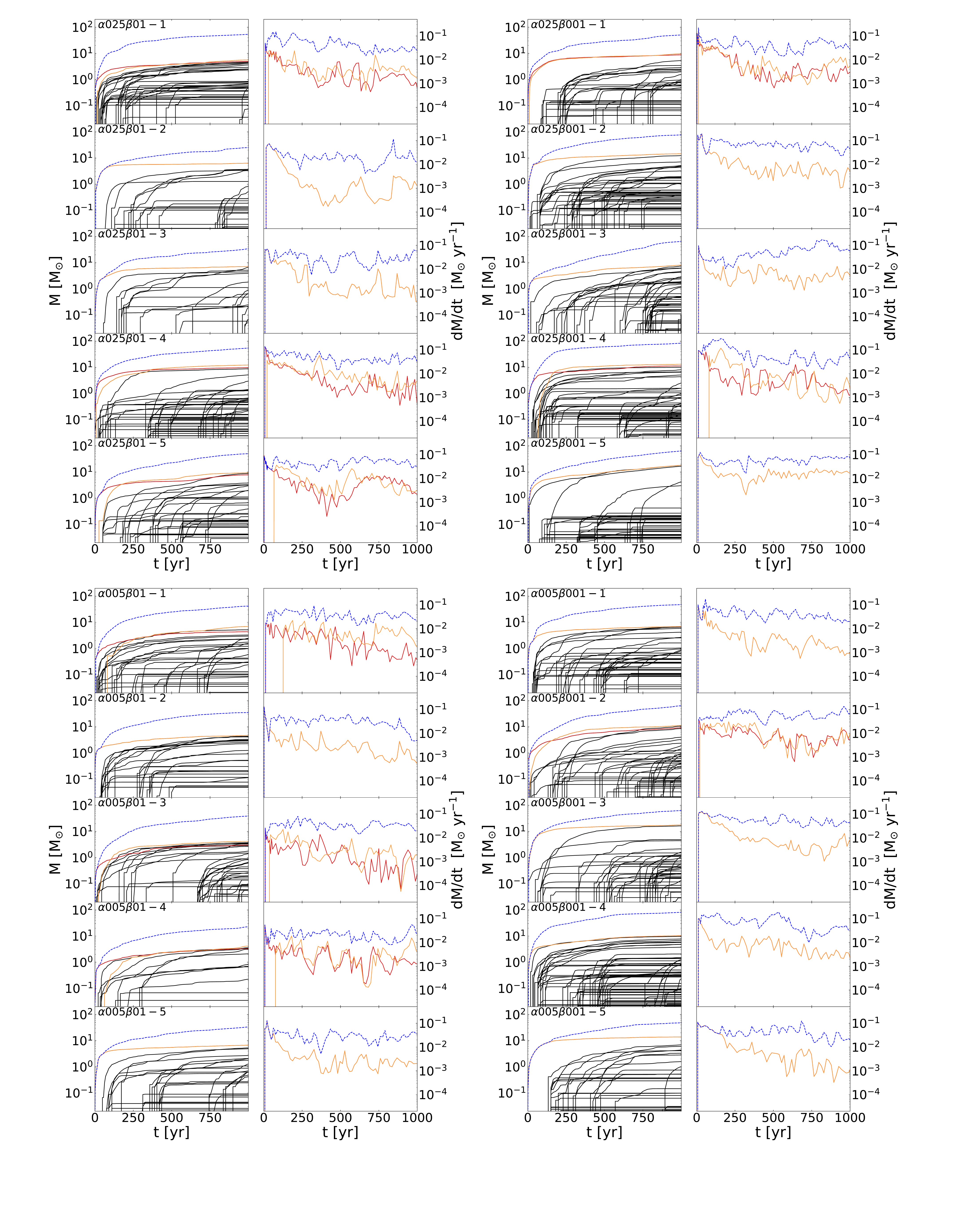} 
\caption{Overview of the mass accretion histories of the mixed runs. The left column in each subplot displays the increase of the total mass in sink (blue dashed line), the mass growth of the first (red) and the most massive sink (orange) together with all other sinks (black). The right column in each subplot shows the cumulative accretion rate (blue dashed line) and the accretion rate of the first (red) and most massive sink (orange). The most massive sink is defined to be the sink with the highest mass at $t \sim 1000 \, {\rm yr}$. As can be seen clearly, the accretion rates are highly time-variable. Nevertheless, the cumulative accretion rate generally remains between $10^{-2}$ and $10^{-1} \, {\rm M_{\odot}} \, {\rm yr^{-1}}$ throughout the period we study, with a weak trend towards lower cumulative accretion rates in simulations with higher levels of rotational energy.
\label{fig:Macc1}} 
\end{figure*}

\begin{figure*}
\centering 
\includegraphics[height=1.2\textwidth]{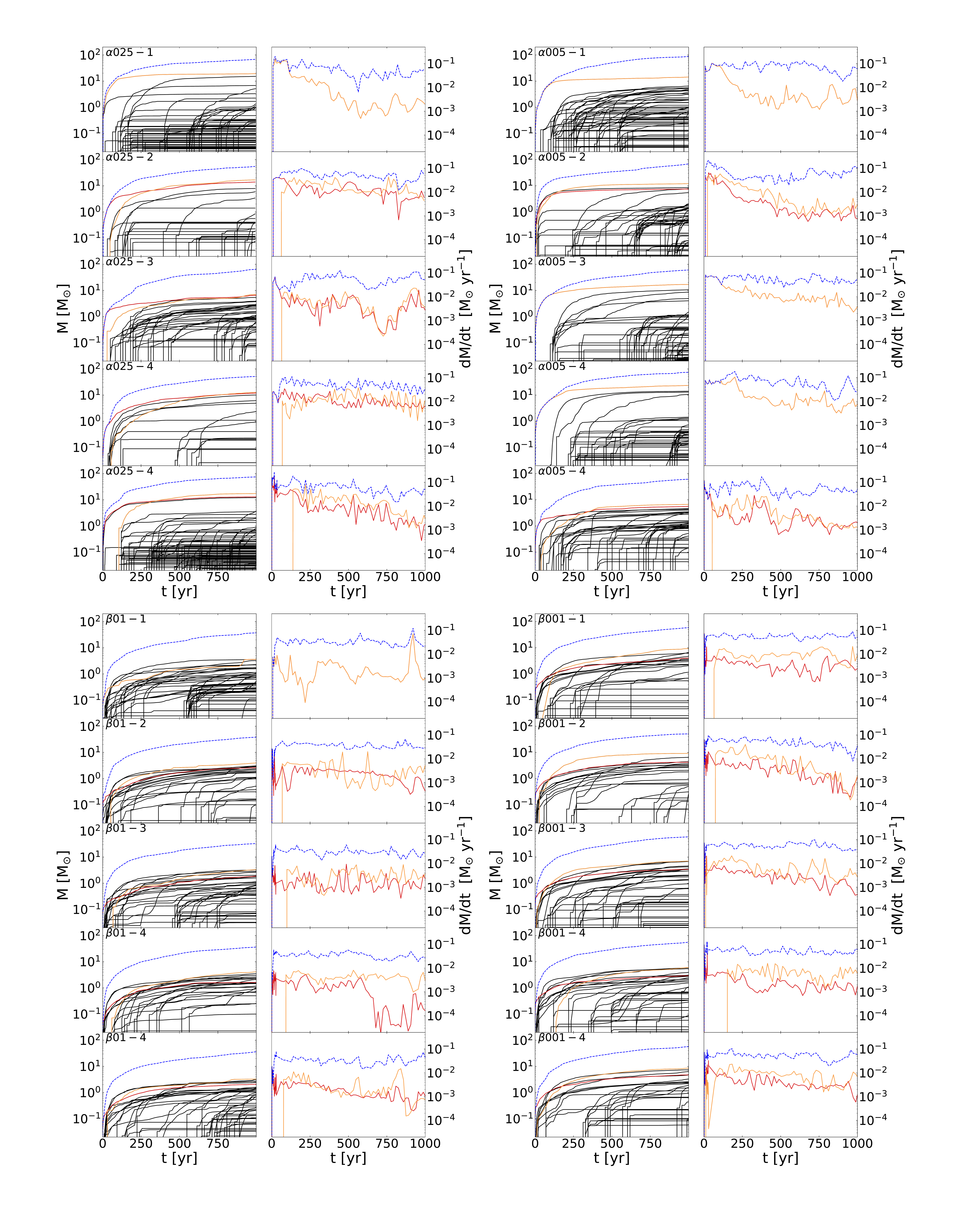} 
\caption{Overview of the mass accretion histories of the purely turbulent (top subplots) and the purely rotational (bottom subplots) runs. The color and line scheme is the same as in Fig. \ref{fig:Macc1} above. Once again, most of the accretion rates are highly time-variable.
\label{fig:Macc2}} 
\end{figure*}

In all of our protostellar systems, we find high cumulative\footnote{We define cumulative accretion rates as the sum of the accretion rates of all sink particles within a simulation.} accretion rates $10^{-2} < \dot{M}_{\mathrm{acc}} < 1 \, \rm M_{\odot} \, yr^{-1}$. Usually, in the first few years after the first sink has formed $\dot{M}_{\mathrm{acc}} \sim 10^{-1} \, \rm M_{\odot} yr^{-1}$, dropping off slowly with time thereafter, although there are a few realizations in which it increases slowly with time. It rarely falls below values of a few times $10^{-2} \, \rm M_{\odot} yr^{-1}$ over the period of time studied in our simulations. Only in one realization, $\alpha005\beta01$-2, does it drop below $10^{-2} \, \rm M_{\odot} yr^{-1}$  before $t \sim 1000 \, \rm yr$. 

In Figures~\ref{fig:Macc1} and \ref{fig:Macc2}, we show the evolution of the individual and total sink masses (left columns) and individual and cumulative accretion rates (right columns) for all realizations and setups. The evolution of the total mass in sinks and the cumulative accretion rate is given by the blue dashed line. Red lines indicate the behavior of the sink which formed first in each simulation, while orange lines indicate the behavior of the most massive sink (defined as the sink with the highest mass at $t=1000 \, \rm yr$). If no red line is visible, this implies that the sink which formed first is also the most massive one.

For all of our runs, the cumulative accretion rate generally remains between $10^{-2} - 10^{-1} \rm M_{\odot} \, yr^{-1}$ throughout the period we study. There is some scatter between the rate in different realizations of a given setup, with stronger and more irregular variations being visible in the runs including turbulence compared to the purely rotational runs. However, overall the difference between different realizations is relatively small, as is also clear from the fact that the total amount of mass accreted by the sinks after 1000 years, $M_{\rm tot}$, varies by at most a factor of two between the different realizations of a given setup. We find a weak trend indicating that runs with higher levels of rotation have lower cumulative accretion rates. This trend can be seen best when comparing the two purely rotational setups with each other. We have already seen the consequences of this behavior in Section~\ref{sec:nummass} where we find an obvious gap in the range of values of $\langle M_{\rm tot} \rangle$ between setups including high level of rotation and the rest of the setups. However, the overall difference between the various setups is still small, consistent with the idea that it is the temperature and density structure of the infalling gas, rather than its level of turbulence or rotation, that is primarily responsible for determining the overall accretion rate.

In the left columns of the subplots of Figures~\ref{fig:Macc1} and \ref{fig:Macc2}, we show how the mass of each individual sink evolves in each simulation. We find that many sinks stop accreting shortly after their formation, while some continue to accrete over a longer period. However, we do not see a clear relationship between when a sink forms and how long it continues to accrete (see \citealt{Schmeja2004} for a similar result in the context of present-day star formation).

In the following we concentrate on describing the evolution of the most massive sink particle and the first sink particle to form. We find that in all realizations and setups, the most massive sink forms before $t=200 \, \rm yr$. The most massive sink after 1000~yr is always one of the first sinks to form, but often not the very first to form (although there are many cases where it is). Initially, the accretion rate onto the first protostar is very large, but as more protostars form it decreases significantly, ending up with a value anywhere between $\sim 10^{-4} \, \rm M_{\odot} yr^{-1}$ and $\sim 10^{-2} \, \rm M_{\odot} yr^{-1}$ after $t=1000 \, \rm yr$. Something similar is true for accretion onto the most massive sink, although in this case the fall-off in the accretion rate is generally less pronounced.  Both rates show much stronger time variability than the cumulative accretion rate, and in some cases short bursts of very rapid accretion are seen (see e.g.\ run $\beta$01-1). Both rates also show considerable variability between the different realizations of a given setup, suggesting that while the turbulence and rotation have only a weak influence on the overall accretion rate, they have a strong influence on {\em which} sinks accrete the gas. 

The values and variability that we find in the accretion rates are in agreement with results from previous numerical studies \citep[e.g.][]{2011aClark, 2011bClark, 2011Greif, 2012Smith, 2014StacyBromm, 2014Susa, 2016Stacy, 2018Riaz}. Directly after its formation, the first sink particle can be expected to have an accretion rate as high as $\gtrsim 10^{-1} \, \rm M_{\odot} yr^{-1}$ \citep{2006Yoshida}. Within the high density environment around the first sinks, new fragments soon form. From Fig. \ref{fig:Nev}, we see that in most realizations, secondary protostars form rapidly after the first. The variability of the accretion rates and their eventual continuous decline stems from the gas dynamics within the protostellar accretion disk where strong gravitational torques redistribute angular momentum and allow more mass to flow to some protostars. Furthermore, it originates from the interactions between the sinks as they compete for accreting material which reduces the accretion onto some of the sinks \citep[e.g.][]{2010Peters, 2011Smith, 2011Greif}.
\citet{2013Susa} find that the majority of the high mass stars ($> 30 \, \rm M_{\odot}$) forming in their simulations are the first protostars to form in their halos. We do not see such a trend, although there is usually at least one realization per setup in which the first sink becomes the most massive sink at $t \sim 1000 \, \rm yr$. However, we caution that we cannot yet draw final conclusions about this as we only cover a short period within the accretion history of a primordial protostellar cluster. 

\subsection{Mass function}
\label{sec:MF}

\begin{figure*}
\centering 
\includegraphics[height=1.2\textwidth]{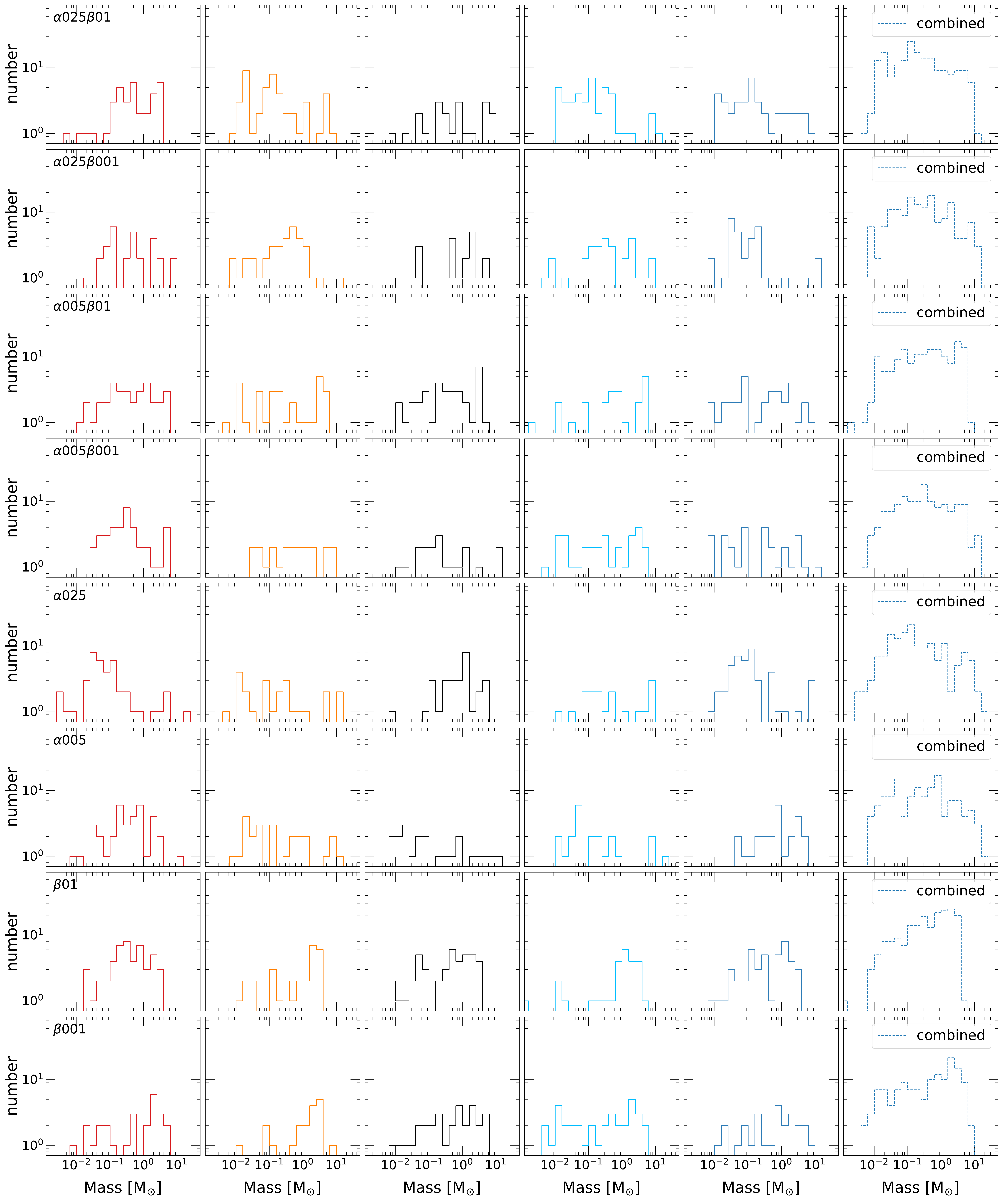} 
\caption{The protostellar mass functions produced in the simulations when the total mass in sinks has reached $\sim 36 \, {\rm M_{\odot}}$ in each realization of each setup. Each row represents a different setup. The first five columns show the mass function for each of the individual realizations. The last column shows the 
combined mass function for each setup, which is derived by summing up the contributions of all individual realizations per setup. The choice of color for the realizations is the same as in Figures~\ref{fig:mev} and \ref{fig:Nev}. The combined mass functions are all relatively flat, although they tend to be somewhat more top-heavy for runs with no initial turbulence. It is also clear that there is considerable variation between the mass functions recovered for different realizations of the same initial setup.
\label{fig:MF}} 
\end{figure*}

In Fig. \ref{fig:MF}, we show the mass function of the sinks when the total mass in sinks has reached $\sim 36 \, {\rm M_{\odot}}$ in each realization of each setup.\footnote{For technical reasons to do with the cadence of the simulation output, we cannot ensure that the mass in sinks in each realization is completely identical, but the variations are at most 1--2\%, as summarized in Table~\ref{tab:turb_result}.} We note that this mass is reached at a different time in each simulation (listed as $t_{\rm same}$ in Table \ref{tab:turb_result}), and that in a few cases it was necessary to continue the simulation beyond $t \sim 1000 \: \rm yr$ in order to reach this value. However, we argue that this enables us to make a more physically meaningful comparison between the different runs than comparing the mass functions at the same elapsed time after the onset of sink particle formation.\footnote{In practice, comparison of the mass functions at $t \sim 1000$~yr yields broadly similar results.}  The particular value of $36 \, {\rm M_{\odot}}$ was chosen for two reasons. First, the fact that most of the simulations have reached this value of $M_{\rm tot}$ by $t \sim 1000 \: {\rm yr}$ means that the computational expense of carrying out the mass function comparison for this value of $M_{\rm tot}$ is not high: only a few runs need continuing to times $t \gg 1000 \: {\rm yr}$. Choosing a significantly larger value of $M_{\rm tot}$ for the comparison would have entailed a much greater additional computational expense. Second, in order for our comparison to be meaningful, we need to carry it out at a point before we have formed many stars with $M \gg 10 \: {\rm M_{\odot}}$, as our lack of stellar feedback renders our simulations unrealistic once many massive stars have formed. As Figure~\ref{fig:MF} demonstrates, at the point when $M_{\rm tot} = 36 \, {\rm M_{\odot}}$, only a very few of the simulations have formed stars with $M > 10 \: {\rm M_{\odot}}$. In the diagrams here, we omit the pure infall models, in which only one sink forms over the course of the simulations. Each row represents a different setup with its individual realizations $1-5$ displayed from left to right in panels one to five. The sixth panel on the right shows the combined mass function for that setup (i.e.\ the sum of the other five mass functions). The color scheme is the same as in Figures \ref{fig:mev} and \ref{fig:Nev}.

We see immediately that there are considerable differences between the mass functions in the different realizations corresponding to a given setup.  Nevertheless, some features of the mass functions remain the same from realization to realization. In most cases, the mass functions are fairly flat, with a large deficit of low mass stars compared to what one would expect from a \citet{1955Salpeter} or \citet{2001Kroupa} mass function. The maximum sink mass is typically $\sim 10 \, {\rm M_{\odot}}$, although as we have already seen, this sink is generally still accreting at a rate of $10^{-3} \: {\rm M_{\odot} \, yr^{-1}}$ or more (see Figures \ref{fig:Macc1} and \ref{fig:Macc2}) and so we would expect to recover larger maximum masses if we were to continue the simulation for longer. The minimum sink mass in each case generally ranges from a few times $10^{-3} \: {\rm M_{\odot}}$ to $10^{-2} \: {\rm M_{\odot}}$. This minimum mass is set by the local Jeans mass in the fragmenting disk, which ranges from $0.004 \lesssim M_{\mathrm{J}} \lesssim 0.01 \, \rm M_{\odot}$ at time of first sink formation. For comparison, the minimum cell mass at the same time is $\lesssim 10^{-6} \, \rm M_{\odot}$ in every simulation, making us confident that the low mass cutoff we see in the mass functions is real and not a numerical artefact. 

Comparing the combined mass functions, we see that although the distributions are fairly flat overall, they are not completely flat. In particular, the runs with a high level of turbulence appear to peak at a mass of around $0.1 \, {\rm M_{\odot}}$, while those with pure rotation peak instead at a mass of a few solar masses. We have checked the similarity of all the combined mass functions with the Kolmogorov-Smirnov (KS) test.  This confirms the visual impression that the combined mass functions from the purely rotational runs are in general not consistent with those in the purely turbulent runs. For reference, we list the KS statistic and corresponding $p$-value for each comparison between setups in Table~\ref{tab:KS-test} in the Appendix. 

Our finding of a flat protostellar mass function is in good agreement with previous studies of fragmentation in Pop.\ III accretion disks that adopted a sink particle-based approach \citep[e.g.][]{2011bClark,2011Greif,2011Smith,2016Stacy}. However, the significant differences we see between the mass functions produced by different realizations of the same setup demonstrates the danger involved when drawing conclusions based on only a small number of simulations of fragmentation, particularly in runs where the gas is highly turbulent. 

It is also important to emphasize that the mass functions shown here are not predictions of the final protostellar IMF. Our simulations only probe a short period in the history of the accretion disk, they do not account for mergers between sinks (see Section~\ref{sec:merger} below), and also they do not account for the effects of radiative feedback, which we expect to play an important role in regulating the high mass portion of the IMF \citep[see e.g.][]{2014Susa,2015Hirano,2016Stacy}. 

\subsection{Stellar encounter and merging}
\label{sec:merger}

\begin{figure*}
\centering 
\includegraphics[height=1.2\textwidth]{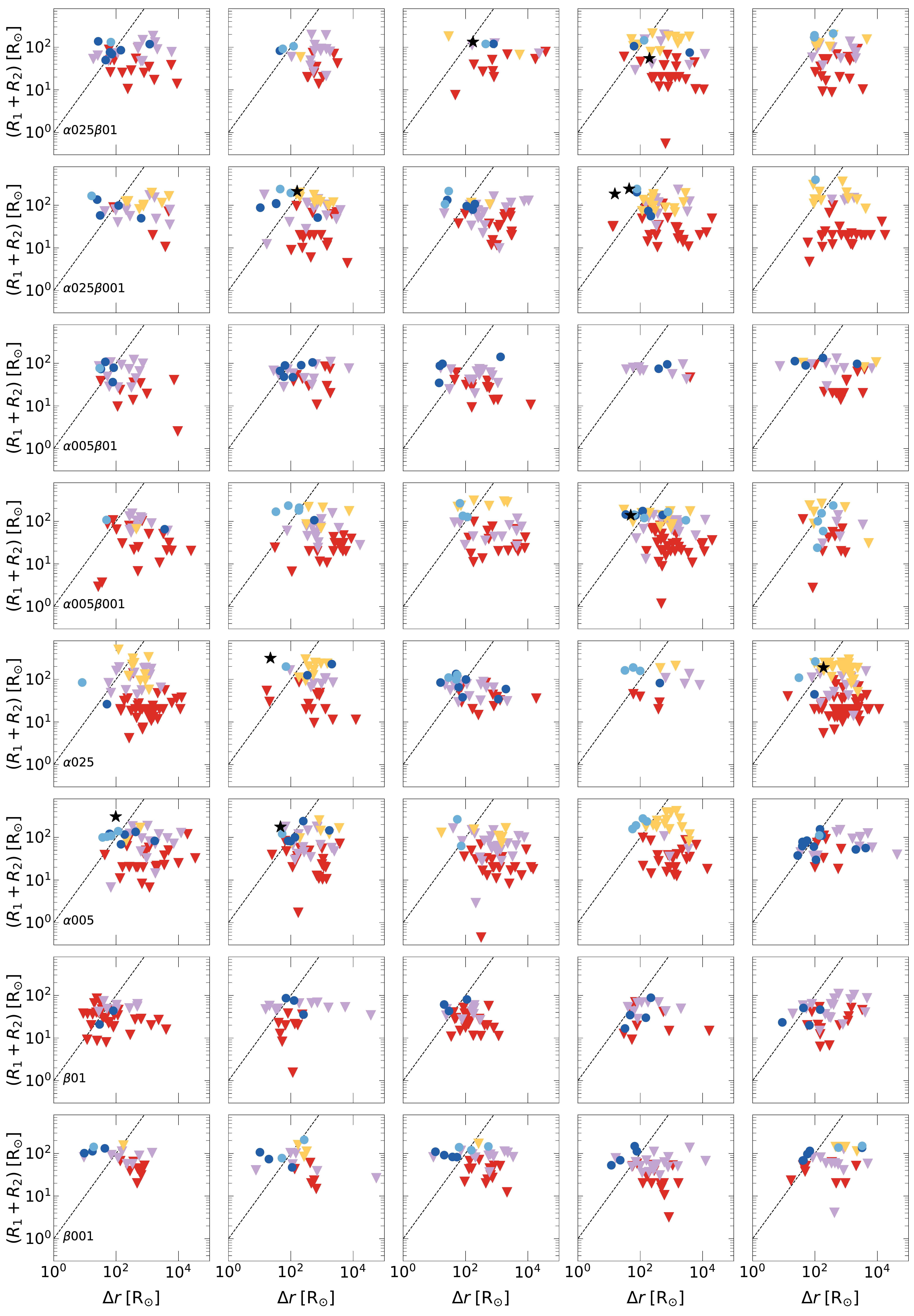} 
\caption{Comparison of the sum of the protostellar radii, $R_{\mathrm{1}} + R_{\mathrm{2}}$, versus the minimum protostellar separation, $\Delta r$, for all close encounters in each simulation. The black dashed line indicates where 
$R_{1} + R_{2} = \Delta r$. Points above and to the left of this line indicate encounters likely to lead to mergers. Downward triangles describe encounters of low-mass protostars ($\leq 0.8 \, \rm M_{\odot}$) with other low-mass protostars (red), with medium-mass protostars ($0.8 < M \leq 5 \, \rm M_{\odot}$; bright violet) and with high-mass protostars ($> 5 \, \rm M_{\odot}$; yellow colored). Filled circles are encounters of two medium-mass protostars (dark blue) or of a medium-mass and a high-mass protostar (bright blue). Black stars indicate encounters of two high-mass protostars. In every simulation, a few encounters occur that are likely to lead to mergers according to this merger criterion.
\label{fig:encounter}} 
\end{figure*}

\begin{figure*}
\centering 
\includegraphics[height=1.2\textwidth]{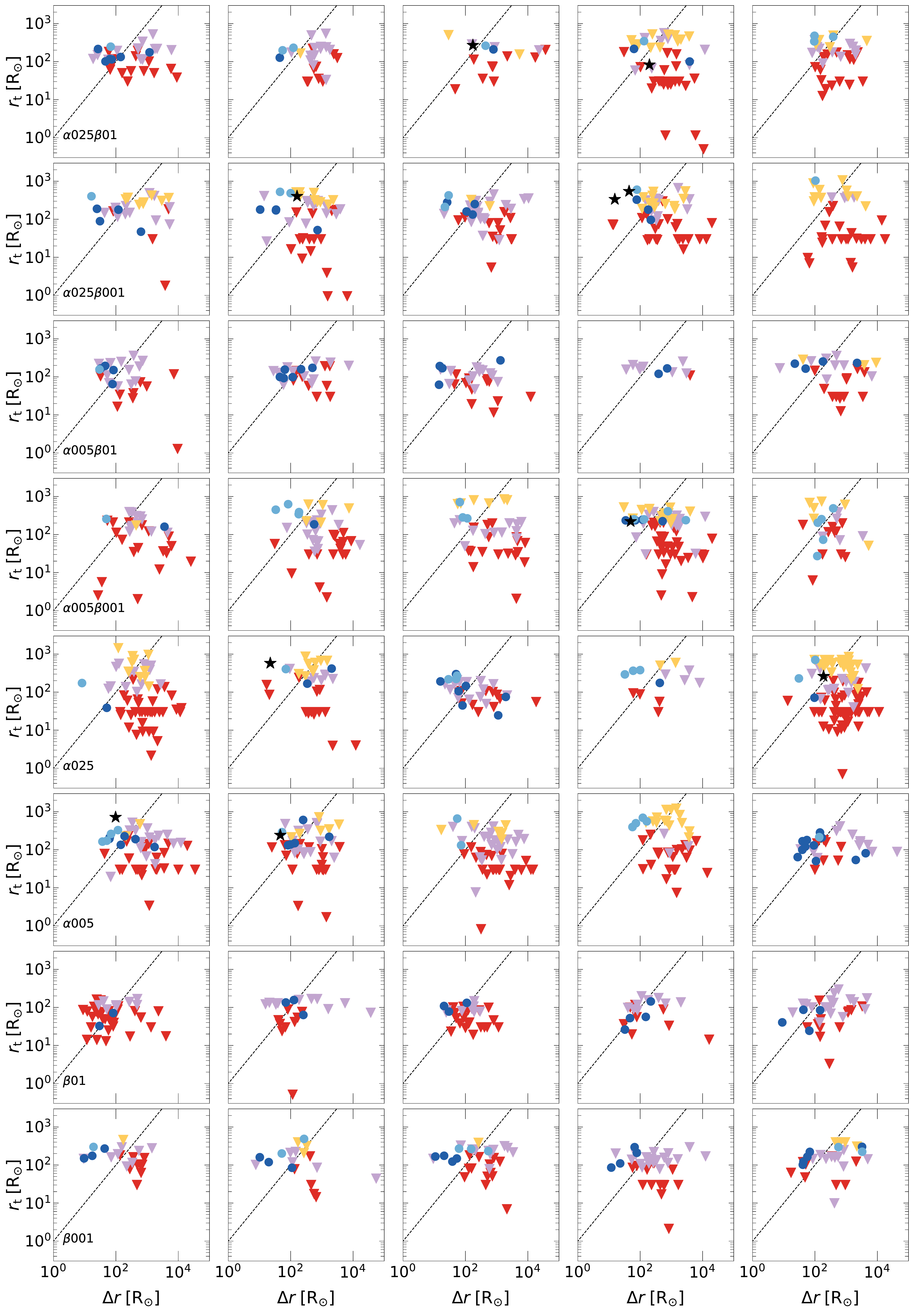} 
\caption{Comparison of the tidal radius $r_{\rm t}$ with the minimum protostellar separation, $\Delta r$, for all close encounters in each simulation. The black dashed line indicates where 
$r_{\rm t} = \Delta r$. The symbols and the color scheme are the same as in  Fig. \ref{fig:encounter}. We see that with the tidal radius merger criterion, a significantly larger fraction of encounters are likely to lead to mergers.
\label{fig:tidal}} 
\end{figure*}

In view of the large number of fragments that form in the gas, and the small size of the region in which they form,  close interactions between protostellar fragments are likely. Previous studies have shown that encounters between protostars are indeed a common feature within Pop III protostellar systems and affect the trajectories of the protostars in question, which might lead to dynamical ejections \citep[e.g.][]{2011Greif, 2011Smith, 2013Susa, 2016Stacy}, formation of binaries or multiple systems \citep[e.g.][]{2010Stacy, 2013Stacy,2018Riaz}, and mergers \citep[e.g.][]{2010Stacy, 2012Greif, 2016Hosokawa, 2016Stacy,Reinoso2018,Boekholt2018,Susa2019}.  We refrain from exploring the properties of binaries or multiple systems since it is likely that they continue to change over the course of the simulation and we cover only a short time in our runs compared to other studies that have tackled this question \citep[e.g.][]{2012Smith, 2013Stacy}. Our current implementation of sink particles within {\sc arepo} does not allow for the possibility of sink merging, so we cannot simulate this self-consistently. Nevertheless, by examining the trajectories of the sink particles, we can study how likely mergers are in our different setups and whether there is any systematic trend with increasing turbulent or rotational energy. 

We consider two kinds of close encounters: ones where the protostellar radii actually touch or overlap (referred to later as the touching-radii scenario) and ones where the lower mass (secondary) protostar approaches to within the tidal radius of the more massive (primary) one (referred to later as the tidal-radius scenario). Numerically, therefore, we compare the distance between the protostars to either the sum of the protostellar radii, $R_{\rm 1} + R_{\rm 2}$, or the tidal radius
\begin{equation}
\label{tidalradius}
r_\mathrm{t} \approx 2.44 R_{\mathrm{1}} \left( \frac{\rho_{\mathrm{1}}}{\rho_{\mathrm{2}}} \right)^{1/3},
\end{equation}
where $R_{\mathrm{1}}$ is the radius of the primary protostar, and $\rho_{\mathrm{1}}$ and $\rho_{\mathrm{2}}$ are the densities of the primary and secondary protostars, respectively. In practice, during the period simulated the protostars will be `fluffy' objects with extended, puffed-up radii \citep{2009Hosokawa, 2010Hosokawa, 2012Smith, 2014Hirano, 2016Hosokawa}, whose densities likely do not vary enormously from protostar to protostar. We therefore follow \citet{2012Smith} and approximate the tidal radius as $r_t \sim 3 R_{\mathrm{1}}$. Finally, we estimate the evolution of the protostellar radii, $R_{\mathrm{1}}$ and $R_{\mathrm{2}}$, by using Eq.~(\ref{eq:radius1}) to~(\ref{eq:radius3}) in Section~\ref{sec:Lacc} of our discussion of the accretion luminosity \citep[see also][]{2011Smith}. Motivated by results from \citet{2012Greif} in which Population III protostar formation was studied without sink particles, we set the initial radius of a newly formed sink to $10 \, \rm R_{\odot}$. We use an accretion rate that is derived by averaging the mass growth over 10 to 20 years. In this way, we smooth out the effects of any rapid increases or decreases in the accretion rate, since more sophisticated protostellar evolution models show that the protostellar radius is more sensitive to the long-term trend of the accretion rate than to its short-term variability \citep{2012Smith}. 

We follow the trajectories of the sinks using the information included in the snapshot files from the simulations. Although these are produced with a high cadence, we nevertheless may miss some close encounters that occur in between snapshots. Our results should therefore be considered as lower limits on the number of close encounters.

In Figs.~\ref{fig:encounter} and \ref{fig:tidal}, we present an overview of the two encounter scenarios for all our setups and realizations. The different symbols and colors indicate which mass range the interacting protostars populate. Downward triangles describe encounters including low-mass protostars ($\leq 0.8 \, \rm M_{\odot}$). Red triangles are encounters of two low-mass protostars, bright violet triangles stand for an encounter between a low-mass and a medium-mass ($0.8 < M \leq 5 \, \rm M_{\odot}$) protostar, and yellow triangles are encounters between a low-mass and a high-mass ($> 5 \, \rm M_{\odot}$) protostar. Filled circles are encounters of two medium-mass protostars (dark blue) or of a medium-mass and a high-mass protostar (bright blue). Black stars indicate encounters of two high-mass protostars. 
The black dashed line describes where the distance between the protostars equals either the sum of their protostellar radii (Fig. \ref{fig:encounter}) or the tidal radius of the primary protostar (Fig. \ref{fig:tidal}). Encounters above and to the left of this line are likely to lead to mergers. We can see immediately that close encounters happen in all realizations for both scenarios, with more encounters occurring in the case of the tidal-radius scenario. This becomes even clearer when we compare columns 3 and 5 in Table \ref{tab:encounter}, where we list the number of merging candidates, $N_{\mathrm{merge}}$ for the two scenarios. The number of merger candidates is always larger in the tidal radius scenario, by anywhere from a factor of two to a factor of several. 

Figures \ref{fig:encounter} \& \ref{fig:tidal} also show that although close encounters take place between stars with a wide range of masses, the minimum separation $\Delta r$ tends to decrease with increasing protostellar mass. Therefore, mergers are far more likely to occur between pairs of protostars that include at least one medium-mass star than between pairs of low-mass protostars. However, mergers between high-mass stars are rare in our simulations, simply because few of these stars form within the period studied here.

\begin{figure*}
\centering 
\includegraphics[width=1.\textwidth]{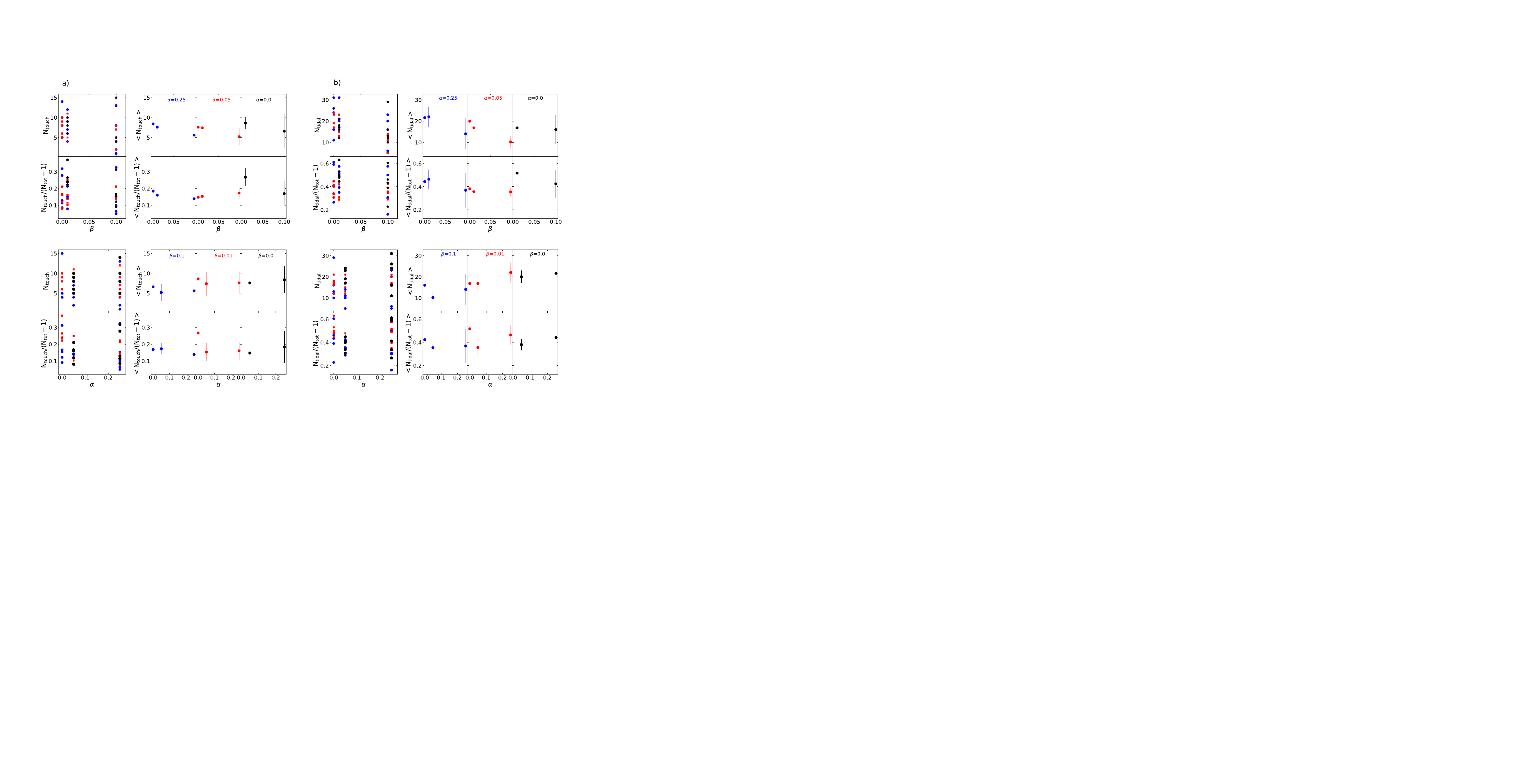} 
\caption{Subplot a): number of mergers per realization in the touching-radii scenario, $N_{\rm touch}$, and the ratio between the number of such mergers per realization and the maximum possible number of mergers per realization, $N_{\rm touch} /(N_{\rm tot} - 1)$, together with their averages. Subplot b) number of mergers in the tidal-radius scenario, $N_{\rm tidal}$, and the ratio between the number of such mergers per realization and the maximum possible number of mergers per realization, $N_{\rm tidal} /(N_{\rm tot} - 1)$, together with their averages. Color-coding in the top diagrams for both a) and b) indicates different values of the turbulent $\alpha$ parameter, while in the bottom diagrams for both a) and b) the color-coding relates to different levels of the rotational $\beta$ parameter.
We see that although the number of mergers varies from setup to setup, this is driven primarily by changes in the total number of sinks. The fraction of the total number of sinks that are involved in mergers shows little dependence on $\alpha$ or $\beta$, with the possible exception of setup $\beta001$.
\label{fig:encounter_average}} 
\end{figure*}

\begin{table}
\footnotesize
\centering
\caption{Overview of the setup averages for the touching radii scenario. From left to right: name of the setup, average number of close encounters per setup, $\langle N_{\mathrm{touch}} \rangle$, and setup average of the ratio between total number of close encounters per realization and maximum possible number of merger per realization, $\langle N_{\mathrm{touch}}/(N_{\mathrm{tot}}-1) \rangle$. The standard deviations in both quantities are also listed.
}
\label{tab:average_encounter1}
\begin{tabular}{lcccc}
\hline
\hline
Setup  & \multicolumn{2}{ |c| }{Touching Radii} \\
                           & $\langle N_{\mathrm{touch}} \rangle$ & $\langle N_{\mathrm{touch}}/(N_{\mathrm{tot}}-1) \rangle$  \\
\midrule

$\beta01$ 		 & 6.6 $\pm$ 4.2 &  0.1692 $\pm$ 0.0763 \\
$\beta001$ 		 & 8.6  $\pm$ 1.5 & 0.2671 $\pm$ 0.0534\\ 
$\alpha025$ 		 & 8.4 $\pm$ 3.4 & 0.1845 $\pm$ 0.0946\\
$\alpha005$ 		 & 7.6 $\pm$ 1.9 & 0.1477 $\pm$ 0.0441 \\
$\alpha025\beta01$   & 5.6 $\pm$ 4.4 &  0.1387 $\pm$ 0.0998 \\
$\alpha025\beta001$ & 7.6 $\pm$ 2.7 &  0.1604 $\pm$ 0.0525 \\ 
$\alpha005\beta01$   & 5.2  $\pm$ 2.1 &  0.1729 $\pm$ 0.0323 \\ 
$\alpha005\beta001$ & 7.4 $\pm$ 3.0 & 0.1534 $\pm$  0.0523 \\ 
\bottomrule
\end{tabular}
\end{table}

\begin{table}
\footnotesize
\centering
\caption{Overview of the setup averages for the tidal-radius scenario. From left to right: name of the setup, average number of close encounters per setup, $\langle N_{\mathrm{tidal}} \rangle$, and setup average of the ratio between total number of close encounters per realization and maximum possible number of merger per realization, $\langle N_{\mathrm{tidal}}/(N_{\mathrm{tot}}-1) \rangle$.  The standard deviations in both quantities are also listed.}
\label{tab:average_encounter2}
\begin{tabular}{lcccccccc}
\hline
\hline
Setup  & \multicolumn{2}{ |c| }{Within $r_{\mathrm{t}}$}\\
                           & $\langle N_{\mathrm{tidal}} \rangle$ &$\langle N_{\mathrm{tidal}}/(N_{\mathrm{tot}}-1) \rangle$ \\
\midrule
$\beta01$ 		 & 16.0 $\pm$ 6.8 & 0.4233 $\pm$ 0.1215\\
$\beta001$ 		 & 16.8 $\pm$ 3.0 & 0.5167 $\pm$ 0.0629\\
$\alpha025$ 		 & 21.6 $\pm$ 7.2 & 0.4431 $\pm$ 0.1370\\
$\alpha005$ 		 & 20.0 $\pm$ 3.0 & 0.3809 $\pm$ 0.0508\\
$\alpha025\beta01$   &14.0 $\pm$ 7.3  & 0.3688 $\pm$ 0.1492 \\
$\alpha025\beta001$ & 22.0 $\pm$ 4.7 & 0.4640 $\pm$ 0.0818\\ 
$\alpha005\beta01$   & 10.2 $\pm$ 3.0 & 0.3541 $\pm$ 0.0429 \\ 
$\alpha005\beta001$ & 16.8 $\pm$ 4.4 & 0.3561 $\pm$ 0.0779\\ 
 \bottomrule
\end{tabular}
\end{table}

In Figure~\ref{fig:encounter_average} we show how the number of mergers and fraction of sinks involved in mergers vary as a function of $\alpha$ and $\beta$ in both the touching radii and the tidal radius scenarios. We see that although the number of mergers varies as we vary $\alpha$ and $\beta$, this is largely driven by the variation in the number of sinks that form: obviously, if a system has more sinks, then more mergers are possible. The fraction of sinks that are involved in mergers does not vary strongly with $\alpha$ or $\beta$ in most cases, remaining constant at around 15\% in the touching radii scenario and around 40\% in the tidal radius scenario \citep[see also][]{Reinoso2018}. The main exception is the $\beta001$ setup: the sinks produced in this setup undergo encounters close enough to lead to mergers at a higher rate than in any of the other setups. However, given the significant variance in the values for each setup, it is unclear whether this represents a real difference or is merely a coincidence.

In summary, we find that close encounters that might lead to mergers occur in all realizations. Depending on what criterion we use for deciding whether two protostars will merge, we find that anywhere between $\sim 10\%$ and $\sim 50 \%$ of protostars may merge, emphasizing that mergers are common in the dense environment in which the protostars form. This result is qualitatively consistent with other studies that find a high level of mergers between Pop.\ III protostars \citep[e.g.][]{2016Stacy,Susa2019}, but our merger fractions are significantly smaller than the values of 60--80\% found in these studies. This is likely a consequence of the different criteria used to decide when protostars merge. For example, \citet{2016Stacy} consider all sinks that come within a single accretion radius of each other to be candidates for merging, as does \citet{Susa2019} in their sink particle run. The critical separation below which sinks may be merged is therefore 1~AU in the simulations of \citeauthor{2016Stacy} and 3~AU in \citeauthor{Susa2019}'s sink particle run. Importantly, this applies {\em for all sinks in the simulation}. For comparison, the critical separation that we have in the tidal-radius scenario is $r_{\rm t} \sim 3 R_{1}$. The most massive and fastest accreting protostars in our simulations have radii of $\sim 100 \: \rm R_{\odot}$, which yields $r_{\rm t} \sim 1.4$~AU. However, most of the protostars have sizes closer to 20--$30 \: \rm R_{\odot}$, yielding much smaller critical separations. Since the merger cross-section scales as the square of the critical separation, this means that most of the protostars formed in our study have cross-sections much smaller than in the other studies, and it is therefore unsurprising that we find fewer mergers. \citet{Susa2019} also presents results from simulations that do not use sink particles, but instead prevent indefinite collapse by artificially stiffening the equation of state above a threshold density $n_{\rm th} = 10^{15} \: {\rm cm^{-3}}$. These runs also find a high degree of fragment merging, but since the physical size of the clumps formed in these runs are comparable to the accretion radius in the corresponding sink particle run, this is once again not surprising. In the end, the most important lesson to take away from this comparison, or indeed from the comparison between our results in the touching-radii and tidal-radii scenarios, is that the amount of merging one recovers in simulations of fragmenting Population III accretion disks is highly sensitive to the method one uses to decide whether or not two fragments should merge. 

\section{Caveats}
\label{sec:caveat}
In the sections above, we have already mentioned a few of the caveats with our study. Here, we discuss some general limitations of our approach that should be addressed in future work.
 
A major uncertainty in our simulations arises due to lack of a sink merging method. Several previous studies have demonstrated that up to $\sim 60\%$ of the sinks undergo encounters that are close enough to plausibly lead to mergers \citep{2011Greif, 2012Greif, 2013StacyBromm, 2016Stacy,Reinoso2018,Boekholt2018}. In our study in Section~\ref{sec:merger}, we also have found that close encounters between two sink particles happen sufficiently often that at least some of these encounters will likely lead to mergers. However, the number of sinks involved in mergers is highly sensitive to the criterion we use to determine whether two sink particles should merge. The two criteria we examine in Section~\ref{sec:merger} -- direct physical collision and tidal capture -- are both over-simplifications compared to the complicated and messy physics of real protostellar mergers. 
Therefore, although our results help to emphasize the importance of treating protostellar mergers in future studies of Pop.\ III star formation, doing so in a physically realistic fashion will not be easy. 

With frequent mergers, the total number of sinks will be reduced, which will subsequently affect the overall dynamical interaction between the remaining sinks in a way which cannot be reliably accounted for by post-processing. Sinks may grow faster in mass, either directly due to mergers, or indirectly due to unimpeded accretion with high accretion rates as fewer sinks compete for the common mass reservoir. This may lead to a change in the shape of the mass function. 

Finally, all of the simulations carried out in this study assume purely hydrodynamical flow and neglect the effects of magnetic fields. This is a common approximation in the study of Pop.\ III star formation, but the accuracy of this approximation is uncertain. Primordial magnetic seed fields can be generated early in the history of the Universe, e.g.\ by early Universe phase transitions \citep{1997Sigl} or the Biermann battery mechanism \citep{1950Biermann, 2008Xu}. These seed fields are extremely weak ($B=10^{-30}$--$10^{-18} \, \rm G$), but can be efficiently amplified to $B \sim 10^{-5} \, \rm G$ or more through the action of the small-scale turbulent dynamo \citep{1968Kazantsev, 2010Schleicher, 2010Sur, 2012Schober} that is driven through turbulent motions arising during the initial collapse of the gas. The impact of this field on the physics of the Pop.\ III accretion disk remains largely unexplored. Ordered magnetic fields could provide efficient magnetic braking, enabling faster inflow of gas from the accretion disk onto the central protostar, and reducing or completely suppressing  fragmentation \citep{2009HennebelleCiardi, 2011Seifried, 2013MachidaDoi, 2014Peters}. However, the field produced by the turbulent dynamo is initially highly disordered, and hence may have much less impact initially. Although our simulations are unable to address this, they do provide a context for assessing whether the fragmentation behavior found in future simulations that do include the magnetic field differs from the hydrodynamical case in a statistically significant fashion, or whether it falls within the range of outcomes found in purely hydrodynamical runs. 

\section{Conclusion}
\label{sec:conclusion}
We have presented a statistical analysis of an ensemble of 3D simulations of Population~III star formation under the influence of different levels of rotation ($\beta=0.01$, $\beta=0.1$) and subsonic turbulence ($\alpha=0.05$, $\alpha=0.25$). The simulations have been performed with the Voronoi moving-mesh code \textsc{arepo} whose quasi-Lagrangian nature makes it an ideal code to study gas collapse. Our version of \textsc{arepo} includes an updated and self-consistent primordial chemistry network together with a treatment for a variable adiabatic index and accretion luminosity heating. We model the collapse of a primordial Bonnor-Ebert sphere beyond the formation of the first protostar and follow the creation of a highly gravitationally unstable protostellar disk system that subsequently fragments and evolves into a Population~III protostellar cluster. Individual protostars that collapse beyond our resolution limit are represented using sink particles. In order to examine the evolution of this cluster, we continue to follow the interactions between the protostars for $1000 \, \rm yr$ after the formation of the first protostar. For each combination of initial conditions, we have run five realizations which vary either in the random number seed used to initialize the turbulent velocity field or the cell configuration of the Voronoi mesh in runs without turbulence.

We find very significant scatter in the results of the individual realizations of setups including turbulence. Our results demonstrate that in order to find general trends in simulations including turbulence, one should always consider a sample of realizations instead of only a single run. The scatter is generally much smaller for the simulations without turbulence, but the numerical noise which is introduced by the variations in the initial cell configuration onto which the Bonnor-Ebert profile is initialized is still enough to produce different fragmentation outcomes.

The main results of the various quantities of our analysis can be summarized as follows. 
\begin{enumerate}
\item Fragmentation of the protostellar disk occurs in all our setups except the {\it pure infall} runs, i.e. those with no initial rotation or turbulence. There, only one protostar forms over the whole course of the simulation. On average, the amount of fragmentation is larger in runs including turbulence. In mixed runs that include both bulk rotation and turbulence, the higher the level of rotation is, the more stable the protostellar disk system becomes against fragmentation. 

\item Within the $1000 \, \rm yr$ period that we model, the cumulative accretion rate generally remains high of the order of $\dot{M} \sim 10^{-2} - 10^{-1} \rm \, M_{\odot} \, yr^{-1}$. There is only a small scatter in the accretion histories of the different setups. Runs including turbulence show a more strongly variable cumulative accretion rate. The overall small difference between the setups is in agreement with the idea that the density and temperature of the infalling material are primarily responsible for the size of the accretion rate rather than the level of turbulence and rotation.

\item The stellar mass function covers a large range of masses from a few $10^{-3} \, \rm M_{\odot}$ to several tens of solar masses for all runs. Its shape is fairly flat for simulations including turbulence, i.e.\ overall indicating top-heavy distributions as already observed in previous studies of Population~III star formation \citep[e.g.][]{2011Greif, 2011bClark, 2014Susa, 2016Stacy}. 
In contrast, the purely rotational runs lead to a completely different distribution with fewer fragments in the low-mass regime and much more towards the high-mass end.   
\end{enumerate}
We note that the exact shape of the mass function would possibly change if we were to account for protostellar mergers, which are likely to be common in these dense clusters. We intend to address this issue in future work. It would also be interesting to explore how the system continues to evolve at times much greater than 1000~yr, but to do this it will be necessary to include a treatment of ionizing and photodissociating radiation from massive Pop.\ III stars, which will start to become important at these times \citep[see e.g.][]{2014Susa, 2015Hirano,2016Stacy,2016Hosokawa}.

\section*{Acknowledgements}
The authors would like to thank Volker Springel and his team for providing the code \textsc{arepo} and for their patience in answering our questions regarding its operation. Particular thanks in this context go to Rainer Weinberger and R\"{u}diger Pakmor. They would also like to thank the anonymous referee for a constructive report that helped to improve the paper.

Support for KMJW's work on this project was provided by the European Research Council under the European Community's Seventh Framework Programme (FP7/2007-2013) via the ERC Advanced Grant `STARLIGHT: Formation of the First Stars' (project number 339177), by the International Max Planck Research School for Astronomy and Cosmic Physics and by the Heidelberg Graduate School of Fundamental Physics. 

SCOG and RSK also acknowledge support from the Deutsche Forschungsgemeinschaft (DFG) via SFB 881 ``The Milky-Way System'' (sub-projects A1, B1, B2 and B8) and also through Germany's Excellence Strategy EXC-2181/1 - 390900948 (the Heidelberg STRUCTURES Excellence Cluster).

The authors gratefully acknowledge support by the state of Baden-W\"{u}rttemberg through bwHPC and the German Research Foundation (DFG) through grant INST 35/1134-1 FUGG. They furthermore acknowledge the data storage service SDS@hd supported by the Ministry of Science, Research and the Arts Baden-W\"{u}rttemberg (MWK) and the German Research Foundation (DFG) through grant INST 35/1314-1 FUGG.

\newpage
\appendix

\section{Kolomogorov-Smirnov statistic}
\begin{table*}
\centering
\caption{Kolmogorov-Smirnov (KS) statistic of the combined sink mass functions (MF). The first line in every field is the KS statistic value and the second line the p-value. A small KS statistic value or a high p-value ($>0.01$, i.e. $>1\%$) indicates that the combined MF distributions of the two compared setups are consistent with being drawn from the same underlying distribution.}
\label{tab:KS-test}
\begin{tabular}{l|l|l|l|l|l|l|l|l|}
                    & $\beta01$  & $\beta001$ & $\alpha025$ & $\alpha005$ & $\alpha025\beta01$ & $\alpha025\beta001$ & $\alpha005\beta01$ & $\alpha005\beta001$ \\ \hline
$\beta01$              	  & 0.00	&0.12	&0.28	&0.16	&0.23	&0.18	&0.11	&0.15 \\
				  &1.00e+00	&1.93e-01	 &1.32e-06	&3.71e-02 	& 8.95e-05	& 8.67e-03	& 1.99e-01	&5.18e-02 \\ \hline
				  
$\beta001$              	& 0.12 &	0.00	&0.28	&0.19	&0.26	&0.22	&0.12	&0.18 \\
				&1.93e-01	&1.00e+00	&1.46e-05&	1.24e-02	&4.23e-05&	1.69e-03&	 2.23e-01	&1.86e-02\\ \hline
				 
$\alpha025$              & 0.28	&0.28	&0.00	&0.15	&0.08	&0.14	&0.20	&0.19\\
				&1.32e-06	 &1.46e-05&	1.00e+00	&7.34e-02 &	6.85e-01& 	7.51e-02& 	2.78e-03& 	1.13e-02\\ \hline
				
$\alpha005$             & 0.16	&0.19	&0.15	&0.00	&0.12	&0.09	&0.11	&0.11\\
				&3.71e-02	 &1.24e-02	&7.34e-02	 &1.00e+00	&2.28e-01 &	6.47e-01	&2.98e-01	 &3.74e-01 \\ \hline
				
$\alpha025\beta01$  & 0.23	&0.26	&0.08	&0.12	&0.00	&0.10	&0.17	&0.13 \\
				&8.95e-05	 &4.23e-05	&6.85e-01 &	2.28e-01	&1.00e+00	&3.83e-01	 &1.71e-02	&1.22e-01 \\ \hline

$\alpha025\beta001$     & 0.18	&0.22	&0.14	&0.09	&0.10	&0.00	&0.12	&0.07\\
				     &8.67e-03	&1.69e-03	 &7.51e-02 &	6.47e-01	& 3.83e-01	& 1.00e+00	&1.93e-01 &	8.58e-01 \\ \hline 
				     
$\alpha005\beta01$    & 0.11	&0.12	&0.20	&0.11	&0.17	&0.12	&0.00	&0.10\\
				  &1.99e-01	&2.23e-01 &	2.78e-03	&2.98e-01 &	1.71e-02	&1.93e-01	 &1.00e+00	&4.44e-01\\ \hline 
				  
$\alpha005\beta001$     & 0.15 &	0.18	&0.19	&0.11	&0.13	&0.07	&0.10	&0.00\\
				     &5.18e-02	&1.86e-02 &	1.13e-02	&3.74e-01 &	1.22e-01	&8.58e-01	 &4.44e-01	&1.00e+00\\ \hline
\end{tabular}
\end{table*}

\section{Encounter tables}

\begin{table*}
\footnotesize
\centering
\caption{Overview of the number of encounters in comparison to total number of sinks at the end of the simulation for each realization. $N_{\mathrm{tot}}$ is the number of sinks after $t \sim 1000 \, \rm yr$. \textit{The full table is available in the online material.}}
\label{tab:encounter}
\begin{tabular}{lccccc}
\hline
\hline
Setup & $N_{\mathrm{tot}}$ & \multicolumn{2}{ |c| }{Touching Radii} & \multicolumn{2}{ |c| }{Within $r_{\mathrm{t}}$}\\
           &                                & $N_{\mathrm{merge}}$ &$N_{\mathrm{merge}}/(N_{\mathrm{tot}}-1)$ & $N_{\mathrm{merge}}$ &$N_{\mathrm{merge}}/(N_{\mathrm{tot}}-1)$\\
\midrule
$\beta01-1$ & 49 & 15 & 0.3125 & 29 & 0.6042\\
$\beta01-2$ & 31 &  5 & 0.1667 & 13 & 0.4333\\
$\beta01-3$ & 42 &  5 & 0.1220 & 16 & 0.3902\\
$\beta01-4$ & 27 &  4  & 0.1538 & 12 & 0.4615\\
$\beta01-5$ & 45 &  4  & 0.0909 & 10 & 0.2273\\
\hline
$\beta001-1$ & 28 & 10 & 0.3704 & 17 & 0.6296\\
$\beta001-2$ & 26 &  6 & 0.2400 & 12 & 0.4800\\
$\beta001-3$ & 37 &  8 & 0.2222 & 16 & 0.4444\\
$\beta001-4$ & 43 & 10 & 0.2381 & 21 & 0.5000\\
$\beta001-5$ & 35 &   9 & 0.2647 & 18 & 0.5294\\
\hline
\end{tabular}
\end{table*}

\begin{table*}
\centering
\caption{Details of the touching-radii scenario. Number of encounters between protostars within the same, $N_{\rm ii}$, or different mass ranges, $N_{\rm ij}$; here indices i and j indicate either low-mass protostars (index `s'; $\leq 0.8 \, \rm M_{\odot}$), medium-mass protostars (index `m'; $0.8 < M \leq 5 \, \rm M_{\odot}$), or high-mass protostars (index `l'; $> 5 \, \rm M_{\odot}$). In addition, we also list the $N_{\rm ii}$ or $N_{\rm ij}$ per total number of measured close encounters in this scenario, i.e. $N_{\rm merge, tot} = N_{\rm ss} + N_{\rm sm} + N_{\rm sl} + N_{\rm mm} + N_{\rm ml} + N_{\rm ll}$.\textit{The full table is available in the online material.}}
\label{appendix:details-touching-radii}
\begin{tabular}{lccccccccccccc}
\hline
\hline
Realization & $N_{\mathrm{ss}}$ & $N_{\mathrm{ss}}$                               & $N_{\mathrm{sm}}$ & $N_{\mathrm{sm}}$           & $N_{\mathrm{sl}}$ & $N_{\mathrm{sl}}$           & $N_{\mathrm{mm}}$ & $N_{\mathrm{mm}}$                               & $N_{\mathrm{ml}}$ & $N_{\mathrm{ml}}$                               & $N_{\mathrm{ll}}$ & $N_{\mathrm{ll}}$                               \\
      &                                     & \multicolumn{1}{r}{$/ N_{\mathrm{merge, tot}}$} &                   & $/ N_{\mathrm{merge, tot}}$ &                   & $/ N_{\mathrm{merge, tot}}$ &                   & \multicolumn{1}{r}{$/ N_{\mathrm{merge, tot}}$} &                   & \multicolumn{1}{r}{$/ N_{\mathrm{merge, tot}}$} &                   & \multicolumn{1}{r}{$/ N_{\mathrm{merge, tot}}$} \\
\midrule
$\beta01-1$ & 11 & 0.7333 & 4 & 0.2667 & 0 & 0.0000 & 0 & 0.0000 & 0 & 0.0000 & 0 & 0.0000 \\
$\beta01-2$ & 0  & 0.0000 & 4 & 0.8000 & 0 & 0.0000 & 1 & 0.2000 & 0 & 0.0000 & 0 & 0.0000 \\
$\beta01-3$ & 0  & 0.0000 & 2 & 0.4000 & 0 & 0.0000 & 3 & 0.6000 & 0 & 0.0000 & 0 & 0.0000 \\
$\beta01-4$ & 2  & 0.5000 & 1 & 0.2500 & 0 & 0.0000 & 1 & 0.2500 & 0 & 0.0000 & 0 & 0.0000 \\
$\beta01-5$ & 0  & 0.0000 & 2 & 0.5000 & 0 & 0.0000 & 2 & 0.5000 & 0 & 0.0000 & 0 & 0.0000 \\
\hline
$\beta001-1$ & 0 & 0.0000 & 2 & 0.2000 & 0 & 0.0000 & 6 & 0.6000 & 2 & 0.2000 & 0 & 0.0000 \\
$\beta001-2$ & 0 & 0.0000 & 2 & 0.3333 & 0 & 0.0000 & 3  & 0.5000 &1 & 0.1667 & 0 & 0.0000 \\
$\beta001-3$ & 0 & 0.0000 & 2 & 0.2500 & 0 & 0.0000 & 5 & 0.6250 & 1 & 0.1250 & 0 & 0.0000 \\
$\beta001-4$ & 3 & 0.3000 & 1 & 0.2000 & 0 & 0.0000 & 4  & 0.4000 & 1 & 0.1000 & 0 & 0.0000 \\
$\beta001-5$ & 4 & 0.4444 & 1 & 0.1111 & 0 & 0.0000 & 4  & 0.4444 & 0 & 0.0000 & 0 & 0.0000 \\
\hline
\end{tabular}
\end{table*}

\begin{table*}
\centering
\caption{Details of the tidal-radius scenario. Number of encounters between protostars within the same, $N_{\rm ii}$, or different mass ranges, $N_{\rm ij}$; here indices i and j indicate either low-mass protostars (index `s'; $\leq 0.8 \, \rm M_{\odot}$), medium-mass protostars (index `m'; $0.8 < M \leq 5 \, \rm M_{\odot}$), or high-mass protostars (index `l'; $> 5 \, \rm M_{\odot}$). In addition, we also list the $N_{\rm ii}$ or $N_{\rm ij}$ per total number of measured close encounters in this scenario, i.e. $N_{\rm merge, tot} = N_{\rm ss} + N_{\rm sm} + N_{\rm sl} + N_{\rm mm} + N_{\rm ml} + N_{\rm ll}$. \textit{The full table is available in the online material.}}
\label{appendix:details-tidal-radius}
\begin{tabular}{lccccccccccccc}
\hline
\hline
Realization & $N_{\mathrm{ss}}$ & $N_{\mathrm{ss}}$                               & $N_{\mathrm{sm}}$ & $N_{\mathrm{sm}}$           & $N_{\mathrm{sl}}$ & $N_{\mathrm{sl}}$           & $N_{\mathrm{mm}}$ & $N_{\mathrm{mm}}$                               & $N_{\mathrm{ml}}$ & $N_{\mathrm{ml}}$                               & $N_{\mathrm{ll}}$ & $N_{\mathrm{ll}}$                               \\
      &                                     & \multicolumn{1}{r}{$/ N_{\mathrm{merge, tot}}$} &                   & $/ N_{\mathrm{merge, tot}}$ &                   & $/ N_{\mathrm{merge, tot}}$ &                   & \multicolumn{1}{r}{$/ N_{\mathrm{merge, tot}}$} &                   & \multicolumn{1}{r}{$/ N_{\mathrm{merge, tot}}$} &                   & \multicolumn{1}{r}{$/ N_{\mathrm{merge, tot}}$} \\
\midrule
$\beta01-1$  & 18 & 0.6207 & 10 & 0.3448 & 0 & 0.0000 & 1  & 0.0345 & 0 & 0.0000 & 0 & 0.0000 \\
$\beta01-2$  & 2  & 0.1538 & 8   & 0.6154 & 0 & 0.0000 & 3  & 0.2308 & 0 & 0.0000 & 0 & 0.0000 \\
$\beta01-3$  & 6  & 0.3750 & 5   & 0.3125 & 0 & 0.0000 & 5 & 0.3125 & 0 & 0.0000 & 0 & 0.0000 \\
$\beta01-4$  & 3  & 0.2500 & 7   & 0.5833 & 0 & 0.0000 & 2  & 0.1667 & 0 & 0.0000 & 0 & 0.0000 \\
$\beta01-5$  & 4  & 0.4000 & 4  & 0.4000 & 0 & 0.0000 & 2  & 0.2000 & 0 & 0.0000 & 0 & 0.0000 \\
\hline
$\beta001-1$  & 2  & 0.1176 & 4  & 0.2353 & 1 & 0.0588 & 8  & 0.4706 & 2 & 0.1176 & 0 & 0.0000 \\
$\beta001-2$  & 0  & 0.0000 & 5  & 0.4167 & 1 & 0.0833 & 4  & 0.3333 & 2 & 0.1667 & 0 & 0.0000 \\
$\beta001-3$  & 1  & 0.0625 & 5  & 0.3125 & 2 & 0.1250 & 6  & 0.3750 & 2 & 0.1250 & 0 & 0.0000 \\
$\beta001-4$  & 6  & 0.2857 & 7  & 0.3333 & 2 & 0.0952 & 5  & 0.2381 & 1 & 0.0476 & 0 & 0.0000 \\
$\beta001-5$  & 7  & 0.3889 & 4  & 0.2222 & 0 & 0.0000 & 7  & 0.3889 & 0 & 0.0000 & 0 & 0.0000 \\
\hline
\end{tabular}
\end{table*}

\bibliography{MNRAS}{}
\bibliographystyle{mnras}

\label{lastpage}
 
\end{document}